\documentclass[aps,pre,superscriptaddress,preprintnumbers,twocolumn,amsmath,amssymb]{revtex4}

\usepackage{graphicx}
\usepackage{dcolumn}
\usepackage{bm}
\usepackage{units}[]
\usepackage[version=4]{mhchem}
\usepackage{color}
\usepackage[caption=false]{subfig}
\usepackage{empheq}

\newcommand{\ee}[0]{\hat{\boldsymbol{e}}}

\newcommand{\ff}[0]{\boldsymbol{f}}

\newcommand{\rr}[0]{\boldsymbol{r}}

\newcommand{\uu}[0]{\boldsymbol{u}}

\newcommand{\beginsupplement}{%
        \setcounter{equation}{0}
        \renewcommand{\theequation}{S\arabic{equation}}%
        \setcounter{figure}{0}
        \renewcommand{\thefigure}{S\arabic{figure}}%
     }

\usepackage{gensymb}    
     
\begin{document}

\title{Experimental Observation of Flow Fields Around Active Janus Spheres}

\author{Andrew I. Campbell}
\affiliation{Department of Chemical and Biological Engineering, University of Sheffield\\Mappin Street, Sheffield S1 3JD, UK}

\author{Stephen J. Ebbens}
 \email{s.ebbens@sheffield.ac.uk}
\affiliation{Department of Chemical and Biological Engineering, University of Sheffield\\Mappin Street, Sheffield S1 3JD, UK}

\author{Pierre Illien}
\affiliation{Sorbonne Universit\'e,  CNRS, Laboratoire PHENIX, 4 place Jussieu, 75005 Paris, France}

\author{Ramin Golestanian}
\email{ramin.golestanian@ds.mpg.de}
\affiliation{Max Planck Institute for Dynamics and Self-Organization (MPIDS), 37077 G\"ottingen, Germany}
\affiliation{Rudolf Peierls Centre for Theoretical Physics, University of Oxford, Oxford OX1 3PU, United Kingdom}

\date{\today}

\begin{abstract}
The hydrodynamic flow field around a catalytically active colloid is probed using particle tracking velocimetry both in the freely swimming state and when kept stationary with an external force. Our measurements provide information about the fluid velocity in the vicinity of the surface of the colloid, and confirm a mechanism for propulsion that was proposed recently. In addition to offering a unified understanding of the nonequilibrium interfacial transport processes at stake, our results open the way to a thorough description of the hydrodynamic interactions between such active particles and understanding their collective dynamics.
\end{abstract}

\maketitle

\paragraph*{Introduction.---}The ability to design active colloids and to accurately control their motion in a fluid environment is one of the cornerstones of modern physical sciences, and has motivated a significant amount of work over the past years \cite{Wong2016,Bechinger2016,Illien2017}. In particular, it has enabled experimental efforts to mimic functions inspired by cellular biology, such as cargo transport or chemical sensing, which might lead to radically new medical applications. From a theoretical point of view, such objects do not obey the laws of equilibrium statistical physics, and the description of their non-equilibrium dynamics remains a major challenge. Understanding these non-equilibrium phenomena could help us develop a new paradigm in engineering by designing emergent behavior.

Phoretic transport has emerged as a prominent mechanism for non-equilibrium activity. It has been known for many decades that the interactions between a particle immersed in a fluid and an inhomogeneous field---which can be a chemical concentration, a temperature, and electrostatic potential---can lead to force-free and torque-free propulsion \cite{Derjaguin1947, Young1959, Anderson1989}. Although the field gradients can be externally imposed on the phoretic particles, a most interesting situation arises when the particle generates them itself. For instance, the particle can bear an ``active site'', which degrades solute particles present in the bulk. The resulting asymmetric distribution of reaction products creates a nonzero slip velocity at the surface of the particle and yields self-propulsion {\cite{Golestanian2005,Golestanian2007,{Reigh2016},{Moran2016}}}.

\begin{figure*}
{\raisebox{40mm}{\large{(a)}}{\includegraphics[width=4.9cm]{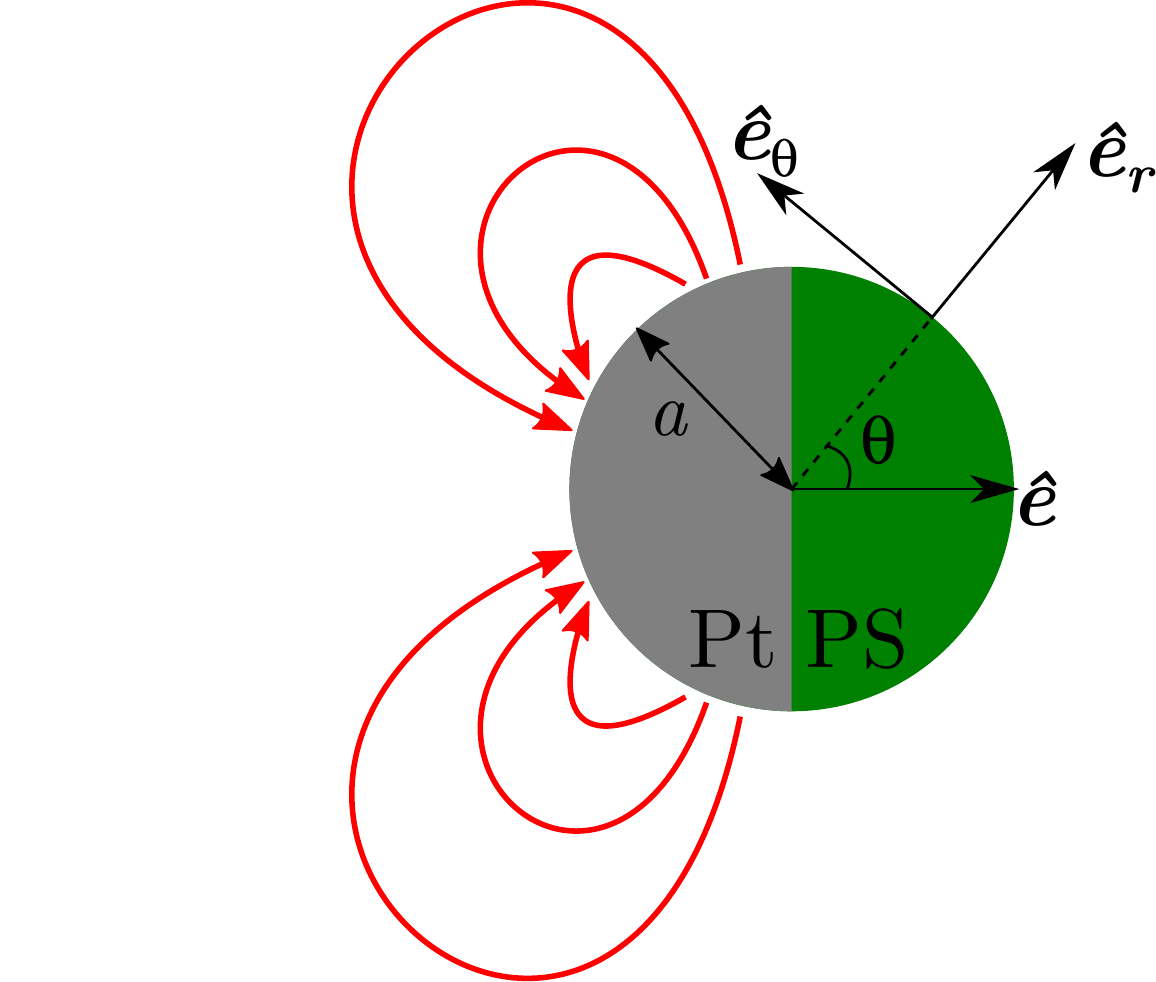}}}
{\raisebox{40mm}{\large{(c)}}{\includegraphics[width=5.5cm]{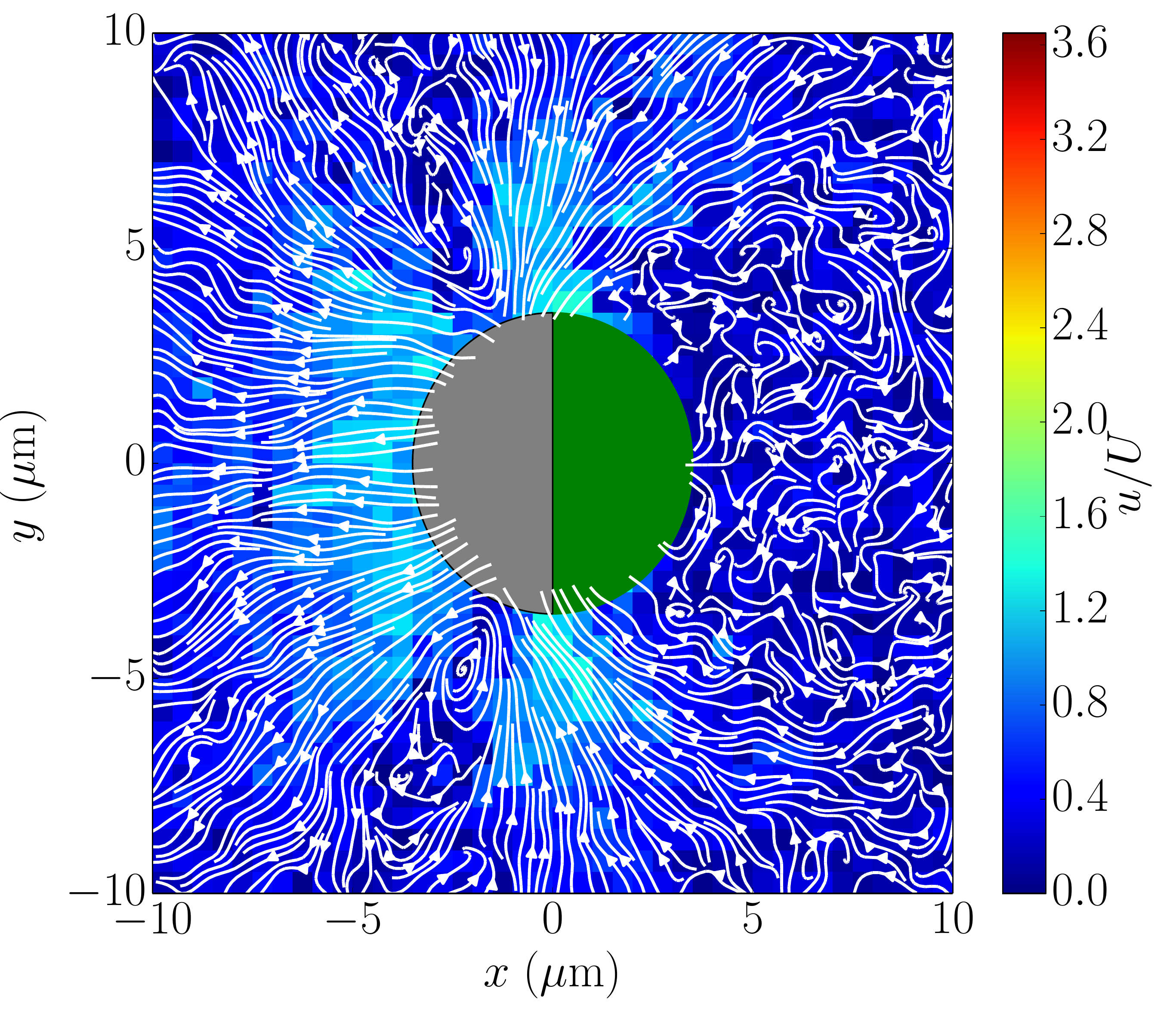}}}
{\raisebox{40mm}{\large{(e)}}{\includegraphics[width=5.5cm]{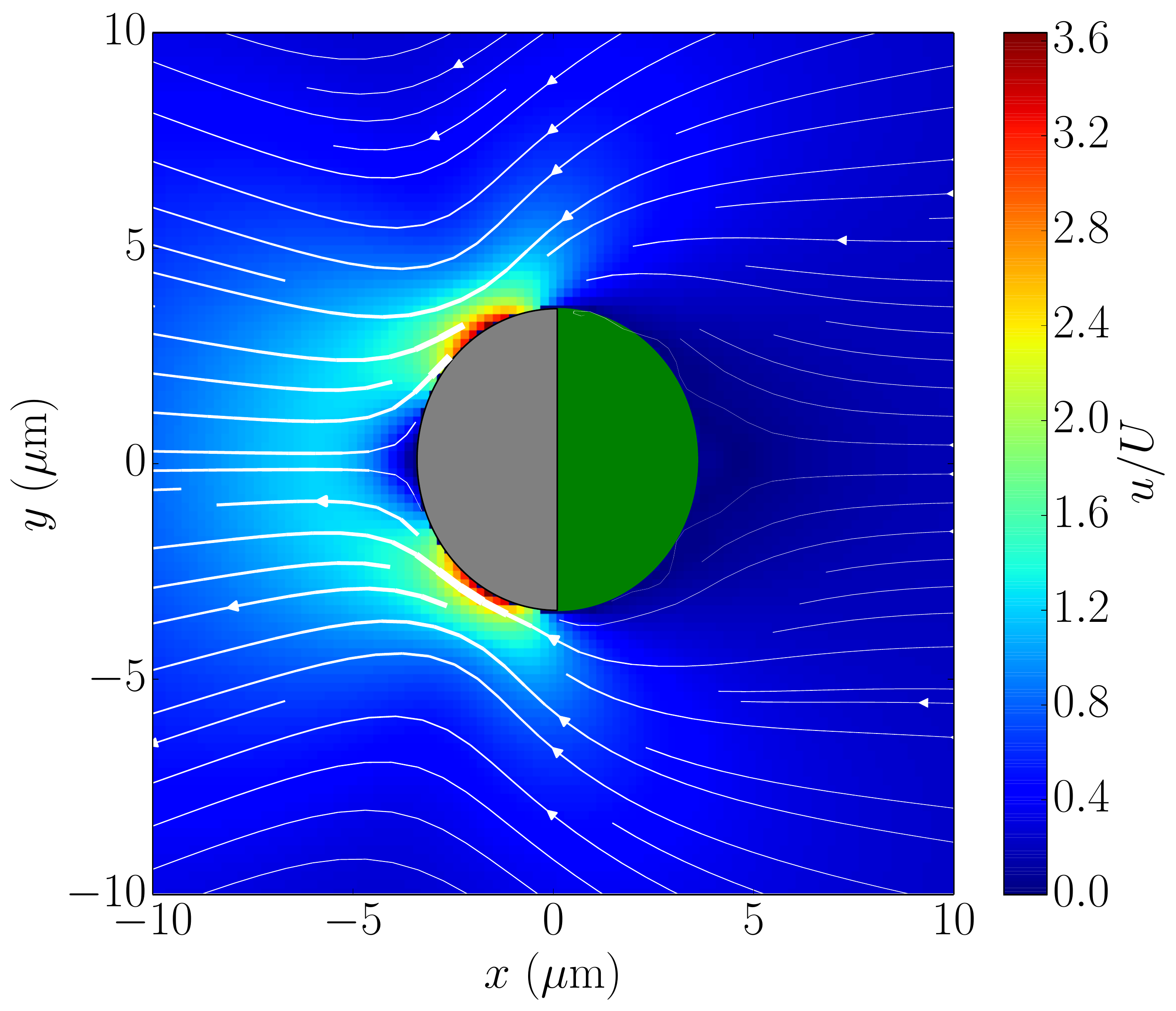}}}
{\raisebox{40mm}{\large{(b)}}{\includegraphics[width=4.8cm]{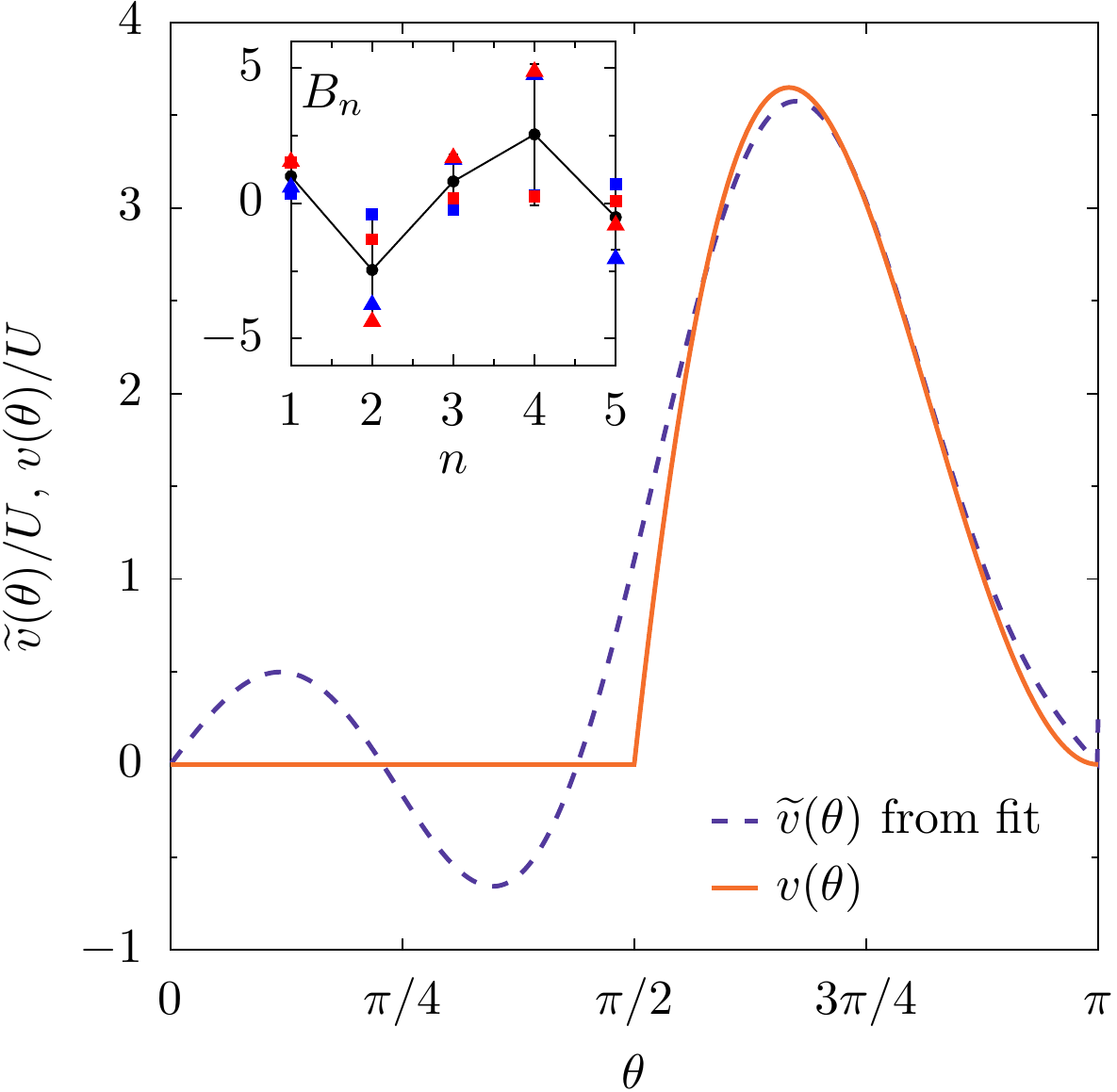}}}
{\raisebox{40mm}{\large{(d)}}{\includegraphics[width=5.5cm]{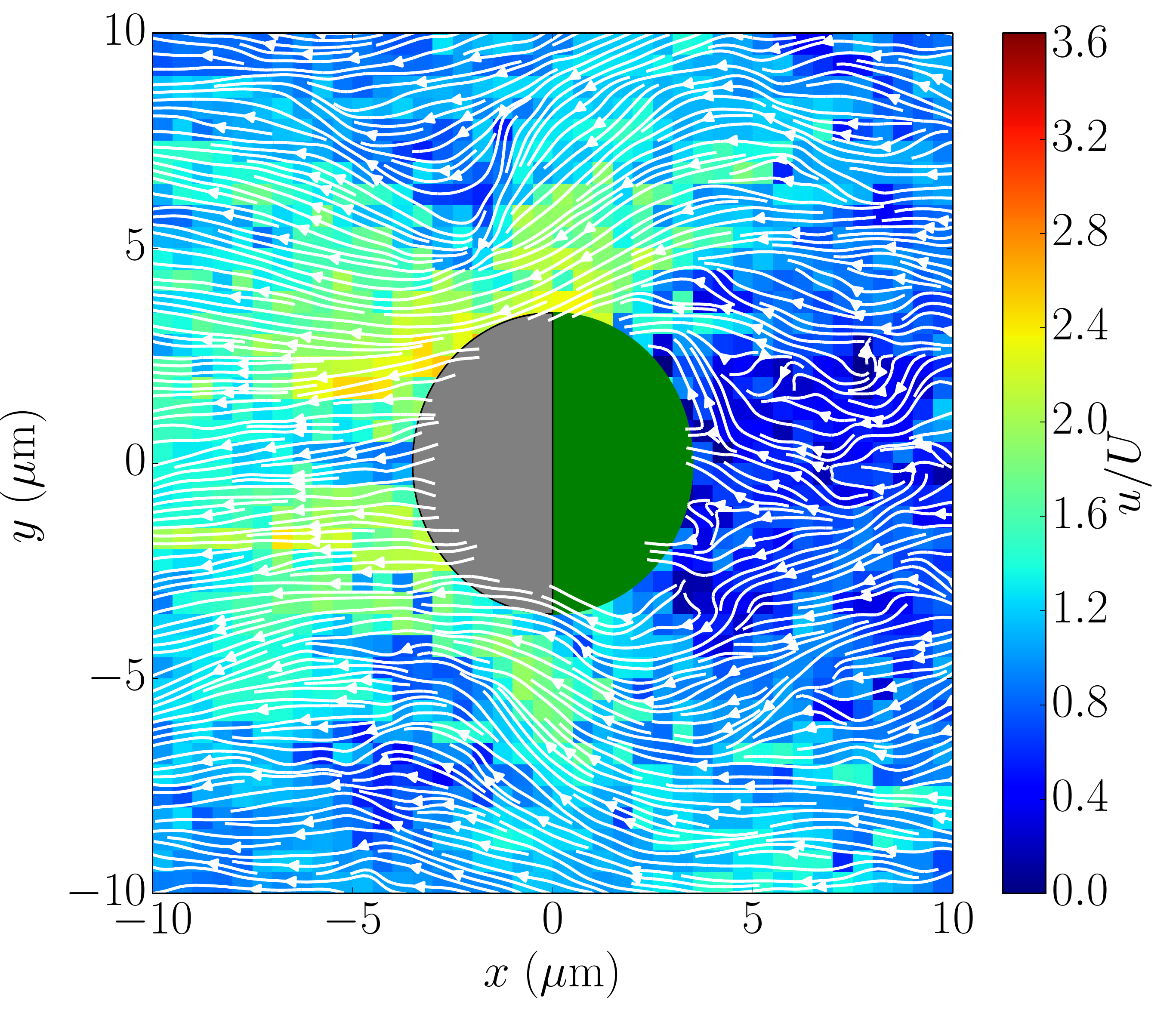}}}
{\raisebox{40mm}{\large{(f)}}{\includegraphics[width=5.5cm]{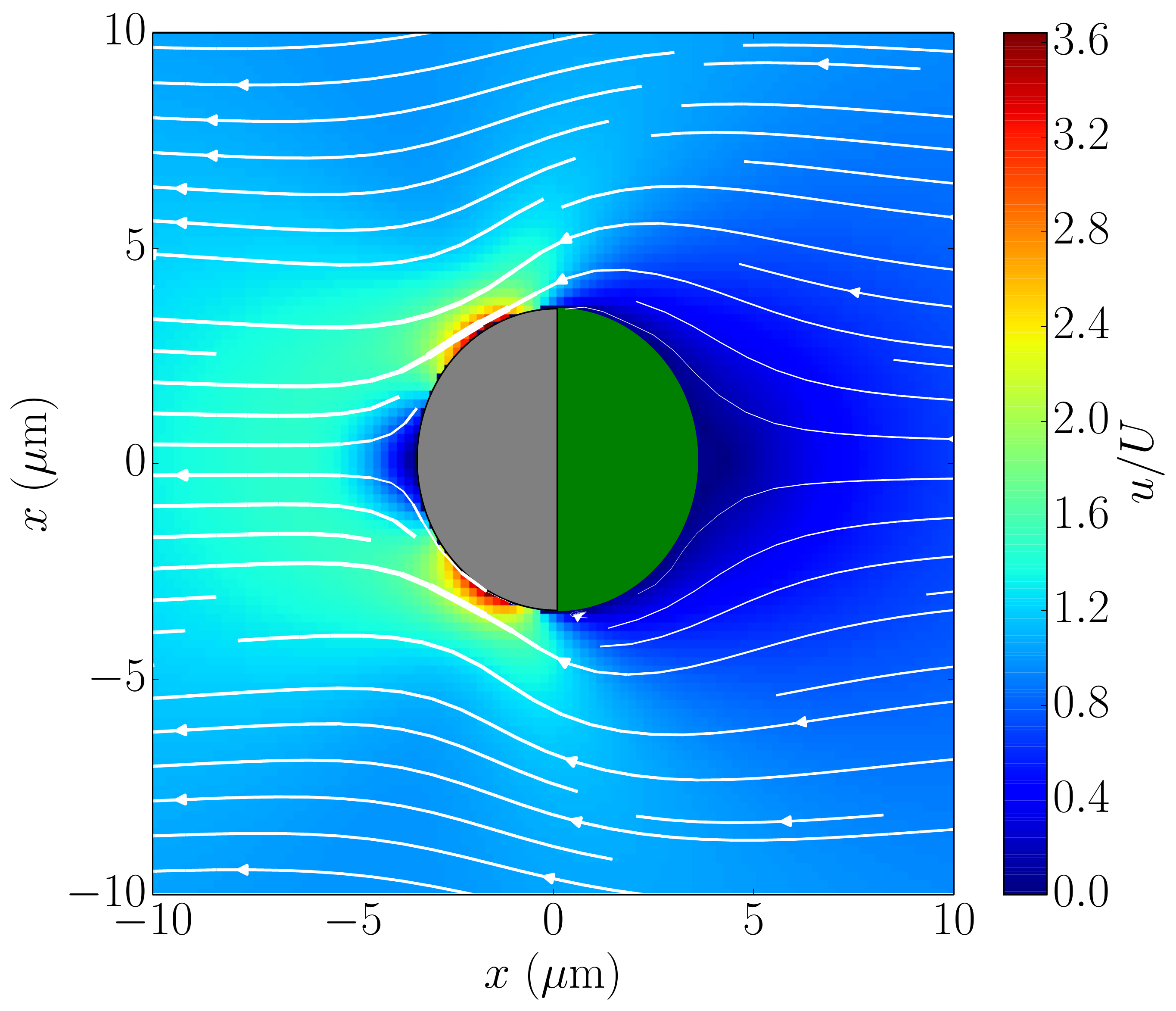}}}
\caption{(a) Sketch of the Pt-PS Janus particle and notations. $\ee$ is the direction of swimming. The red arrows represent the current loops that start at the equator of the particle and end at its pole \cite{Ebbens2014,Ibrahim2017,Das2015}. {(b) Inset: First five coefficients in the expansion of $v(\theta)$ in Legendre polynomials, obtained  as fitting parameters of the experimental radial and tangential velocities, in the cases where the colloid is stuck or freely moving. The averages are shown as black dots, and the error bars correspond to the standard deviations. Main: Plot of the slip velocity $\widetilde{v}(\theta)$ reconstructed from the averages of the Legendre coefficients obtained from the fit (dashed) compared with the simplified expression for $v(\theta)$ we use for further analysis.} (c)-(f) Streamlines around the Janus particles obtained experimentally (left) and analytically (right), in the two situations where the particle is stuck (top) and freely moving (bottom). The background colors represent the magnitude of the velocity $u=|\uu|$ rescaled by the swimming velocity of the Janus particle when it is freely moving.}
\label{streamlines}
\end{figure*}

A number of different particles that break fore-aft symmetry have been designed, including bimetallic rods \cite{Paxton2004} and insulator spheres with only one metal-coated hemisphere \cite{Howse2007} that are able to decompose hydrogen peroxide with Pt as a catalyst. It was shown that bimetallic swimmers are able to decompose fuel simultaneously at both ends. Such an electrochemical process results in an electron transfer across the object, and the associated proton movement in the solution permits self-electrophoretic propulsion. However, despite extensive studies, the phoretic mechanisms at stake in Janus sphere propulsion are not yet well established, and have been the subject of recent debates. Although their constitutive material (typically polystyrene) is an insulator, recent experimental investigations have revealed that electrokinetic effects need to be taken into account together with diffusiophoretic effects to fully describe the propulsion mechanism \cite{Ebbens2014,Brown2014,Brown2017,Ibrahim2017}. The emergence of such effects is permitted by the inhomogeneity of Pt coating, which is thicker at the pole of the colloid than at its equator. We recently proposed a mechanism where the motility of the Pt-PS sphere is linked to closed current loops, which start at the equator of the particle and end at its pole [Fig. \ref{streamlines}(a)]. These conjectured closed proton loops are expected to yield a strongly asymmetric slip velocity across the surface of the colloid \cite{Ebbens2014,Ibrahim2017}. This proposal allowed us to explain the ionic strength sensitivity of the swimming velocity and the non-equilibrium surface alignment phenomenon observed in our experiments \cite{Das2015}. However, direct experimental evidence that proves this conjecture has been lacking so far.


Here, we measure the flow fields around moving Pt-PS Janus particles using Particle Tracking Velocimetry (PTV). {This approach has previously provided insight in to flow fields around stagnant Janus colloids actuated by AC electric fields \cite{Peng2014,Boymelgreen2015}. However, extending this method to our chemically driven system is} comparatively more challenging due to the random stochastic Brownian re-orientation of the moving Pt-PS Janus particles.  
In a related study, the flow field around swimming microorganisms was reported, using photo-tactic guidance to prevent rotation in one case, and also by analysis of tracer particle motion near freely moving cells subject to rotation  \cite{Drescher2010}.  In absence of a mechanism to prevent Pt-PS Janus particle rotation, here we developed  pattern matching image analysis algorithms to allow tracer particle motion to be quantified relative to the frame of reference of a freely moving and rotating Pt-PS Janus particle. Control experiments conducted near a symmetrical catalytically active Pt colloid showed that the tracer motion is dominated by fluid flow rather than chemical gradients.
This method consequently allowed us to arrive at average flow fields derived from observations of many different Pt-PS Janus particles.

\paragraph*{Experimental methods.---}Our experimental system consisted of a \unit[4]{cm} $\times$ \unit[1]{cm} $\times$ \unit[0.1]{cm} glass cuvette filled with a very dilute suspension of $a=\unit[3.5]{\mu\mathrm{m}}$ polystyrene Janus spheres in \unit[10]{wt\%} \ce{H2O2} solution {(no added electrolyte)} \cite{Howse2007}. The Janus spheres are coated on one hemisphere with a thin evaporated layer of \ce{Pt} metal with a maximum thickness of \unit[10]{nm}. To the suspension we added $\unit[3]{\mu\textrm{L}}$ of a \unit[1.0]{wt\%} dispersion of green fluorescent polystyrene spheres ($\textrm{radius}=\unit[0.25]{\mu\mathrm{m}}$) to act as tracer particles. After vigorous shaking to fully disperse the tracer particles the cuvette was placed on the stage of a Nikon Eclipse LV100 microscope fitted with a Nikon 20$\times$ 0.45 N.A. objective. Operating in fluorescence mode using the blue excitation band of a Nikon B2A filter cube the tracer particles at the top of the cuvette were brought into focus.  The calculated depth of field for this optical arrangement is 3.1 $\mu$m, and the focal plane is manually positioned to capture the motion of the tracer colloids at a depth corresponding to the equatorial region of the Janus colloid.  While some tracer particles that are nearer to the substrate will also be in focus, these rapidly move either out of focus and so do not contribute significantly to the flow field, or back towards the equatorial region. After a few minutes equilibration time some of the gravitactic Janus sphere swimmer particles \cite{Campbell2013} had traveled to the top of the cuvette and could be seen translating parallel to the glass surface. After a period some of the Janus spheres became stuck to the surface, probably due to impurities on the surface.

We recorded images (512 $\times$ 512 pixels) of the Janus spheres moving through the field of tracer particles using an Andor Neo camera at a frequency of \unit[100]{Hz}. We note that focussed at the top of the cuvette the bulk of the tracer particles are not in focus. This provided a background level of illumination that enabled us to image the non-fluorescent Janus spheres, where they appeared as a transparent disc with a shadow covering one half of the disc indicating the position and rotation of the \ce{Pt} cap. {For both the moving and stuck colloids the Janus particle appeared as a ``half-moon'' shape consistent with the near surface alignment effect reported in \cite{Das2015}.  This was confirmed by image analysis that found the polar angle to be close to 90\degree. The observation that the static colloids also showed this alignment relative to the substrate, suggests they were translating along the surface before becoming stuck.} Tracking of the tracer particles has been described before \cite{Crocke1996,Howse2007}  but the Janus spheres required a different approach. We used the pattern matching functions of the LabView graphical programming language to obtain the center position of a Janus sphere and the angle of rotation of its \ce{Pt} cap. Tracer particle and Janus sphere trajectories were smoothed to suppress Brownian noise. We then calculated the frame-by-frame vectors along the tracer particle trajectories relative to the position and angle of the Janus sphere. The vectors were then placed on a \unit[0.5]{$\mu$m} grid with the Janus sphere at the origin and an angle of rotation of \unit[0]{$^\textrm{o}$}. Where multiple vectors occupied the same grid position an average vector was calculated. 
1240 s of video data of twenty-five Janus spheres and 1260 s of video data of twenty-one Janus spheres were used to generate the average flow field around the translating and stuck Janus spheres respectively.
Figure \ref{streamlines} shows the experimentally determined flow fields around both stuck and translating Janus spheres, with streamlines generated using the streamplot function of the Python Matplotlib library.

{To assess phoretic effects in a closely related system that is expected to only produce a chemical gradient, and no flow field, $a=2.5 \mu\text{m}$ polystyrene particles were symmetrically coated in Pt metal.  This was achieved using the solution phase salt reduction method that has been previously described, performed without the Pickering emulsion mask step, resulting in a uniformly Pt coated colloid  \cite{Archer2018}.  Tracer motion was again observed and quantified as described above both in the presence and absence of 10 wt \% $\text{H}_2\text{O}_2$.}

\paragraph*{Theoretical description.---}In the low Reynolds number limit, the flow field $\uu$ around the Janus colloid is the solution of the  Stokes problem:
\begin{equation}
\begin{aligned}
 - \eta \nabla^2\uu &= -\nabla p, \\
  \nabla\cdot \uu &= 0,
\end{aligned}
\label{Stokes}
\end{equation}
with boundary conditions imposed by the phoretic mechanisms taking place at the surface of the colloid. 

We first present the solution of these equations in free space, neglecting the nearby solid wall at the top of the experimental cell (see below for a discussion regarding this approximation). The phoretic contribution to the total flow field $\uu^\text{ph}$ originates from a non-zero tangential slip velocity at the surface of the colloid that arises from the current loops \cite{Ebbens2014,Ibrahim2017}, denoted as $v(\theta) \equiv \uu^\text{ph}(r=a,\theta)\cdot\ee_\theta$ for simplicity. The slip velocity $v(\theta)$ is written as an expansion on Legendre polynomials $P_n$:
\begin{equation}
v(\theta)=\sum_{n=1}^\infty B_n V_n(\cos\theta)=\sum_{n=1}^\infty B_n \cdot \frac{2}{n(n+1)}\sin\theta\,P'_n(\cos \theta)
\end{equation}
The coefficients $B_n$ (which have the dimension of a velocity) are not known \textit{a priori} and depend on the details of the phoretic mechanisms taking place at the surface of the colloid.

In the situation where the Janus swimmer is stuck to the glass slide, the total flow field may be written as $\uu=\uu^\text{ph}+\uu^\text{mono}$, where $\uu^\text{ph}$ is the phoretic contribution and $\uu^\text{mono}$ is the contribution from the force monopole that maintains the colloid steady in the laboratory frame of reference.  The contribution from the force monopole $\boldsymbol{f} = -f \ee$ that holds the colloid to the glass slide takes the simple form
\begin{eqnarray}
\label{ }
\uu^{\text{mono}} &=& \frac{1}{2} \frac{f}{6\pi\eta a}  \left[ \left(\frac{a}{r}\right)^3 -3 \left(\frac{a}{r}\right) \right] \frac{\ee \cdot \rr}{r}  \frac{\rr}{r} \nonumber\\
&+&\frac{1}{4} \frac{f}{6\pi\eta a}  \left[ \left(\frac{a}{r}\right)^3 +3 \left(\frac{a}{r}\right) \right] \left(\frac{\ee \cdot \rr}{r}  \frac{\rr}{r} - \ee \right),
\end{eqnarray}
where $\ee$ is the direction of swimming. The value of the force $f$ is computed by enforcing the steadiness of the colloid in the laboratory frame of reference. The flow field around the colloid when it is freely moving is obtained in a similar fashion, but without the contribution from the force monopole and subject to the boundary condition $\lim_{r\to\infty} \uu = -U\ee$, where $U=2B_1/3$ is the swimming velocity of the colloid. In this experiment, the average velocity for the moving Janus colloids determined by mean-squared displacement fitting was $U=5.37 \mu\text{m} \pm 1.64 \mu\text{m}$, and this value was used to normalize the presented data.

In the Supplementary Information \cite{SI}, we give the expression of the velocity field around the stuck colloid  $\uu_\text{s}(r,\theta)$ and around a freely moving colloid  $\uu_\text{m}(r,\theta)$ in the equatorial plane. They only depend on the sets of Legendre coefficients $\{B_n\}_{n\geq 1}$. Each of these two fields has components along $\ee_r$ and $\ee_\theta$, and are measured experimentally. We therefore with get four data sets, that depend on the two variables $(r,\theta)$ and the parameters $\{B_n\}$. We assume that the sums over the Legendre polynomials are truncated at some order $N$ (here we use $N=5$). We use the functions determined analytically \cite{SI} as fitting functions for the experimental data, so that we obtain four sets of fit parameters $B_1,\dots,B_5$. The data sets are shown in the inset of Fig. \ref{streamlines}(b), together with the average and the standard deviation for each case. On this Figure, we also show the slip velocity $\widetilde{v}(\theta) = \sum_{n=1}^5 B_n V_n(\cos\theta)$ reconstructed from these fit parameters. We observe that it is greater in magnitude on the Pt hemisphere, very close to zero at $\theta=\pi/2$ and $\pi$, and has a maximum between $\pi/2$ and $\pi$.

\textit{A simplified slip velocity function.---} 
For further applications, it can be useful to use and can be chosen to have the simplified form \cite{Das2015}
\begin{equation}
\label{slip}
v(\theta) =
\begin{cases}
  v_0(1+\cos\theta)(-\cos\theta)     & \text{ for $\pi/2<\theta<\pi$}, \\
  0    & \text{otherwise}.
\end{cases}
\end{equation}
[See Fig. \ref{streamlines}(a) for a definition of the notations, and Fig. \ref{streamlines}(b) for a representation of the slip velocity as a function of the polar angle $\theta$]. 
Taking the expression of the slip velocity given in Eq. \eqref{slip}, we find that the swimming velocity $U$ is related to the parameter $v_0$ through the relation $U =v_0/K$ where $K =1/\left(\frac{1}{6} - \frac{\pi}{32} \right) \simeq 14.6$.
Finally, the expressions of the velocity field in the two situations where the particle is stuck and freely moving can be written explicitly as Legendre polynomials expansions \cite{SI}.\\

\begin{figure}
\begin{center}
\includegraphics[width=0.9\columnwidth]{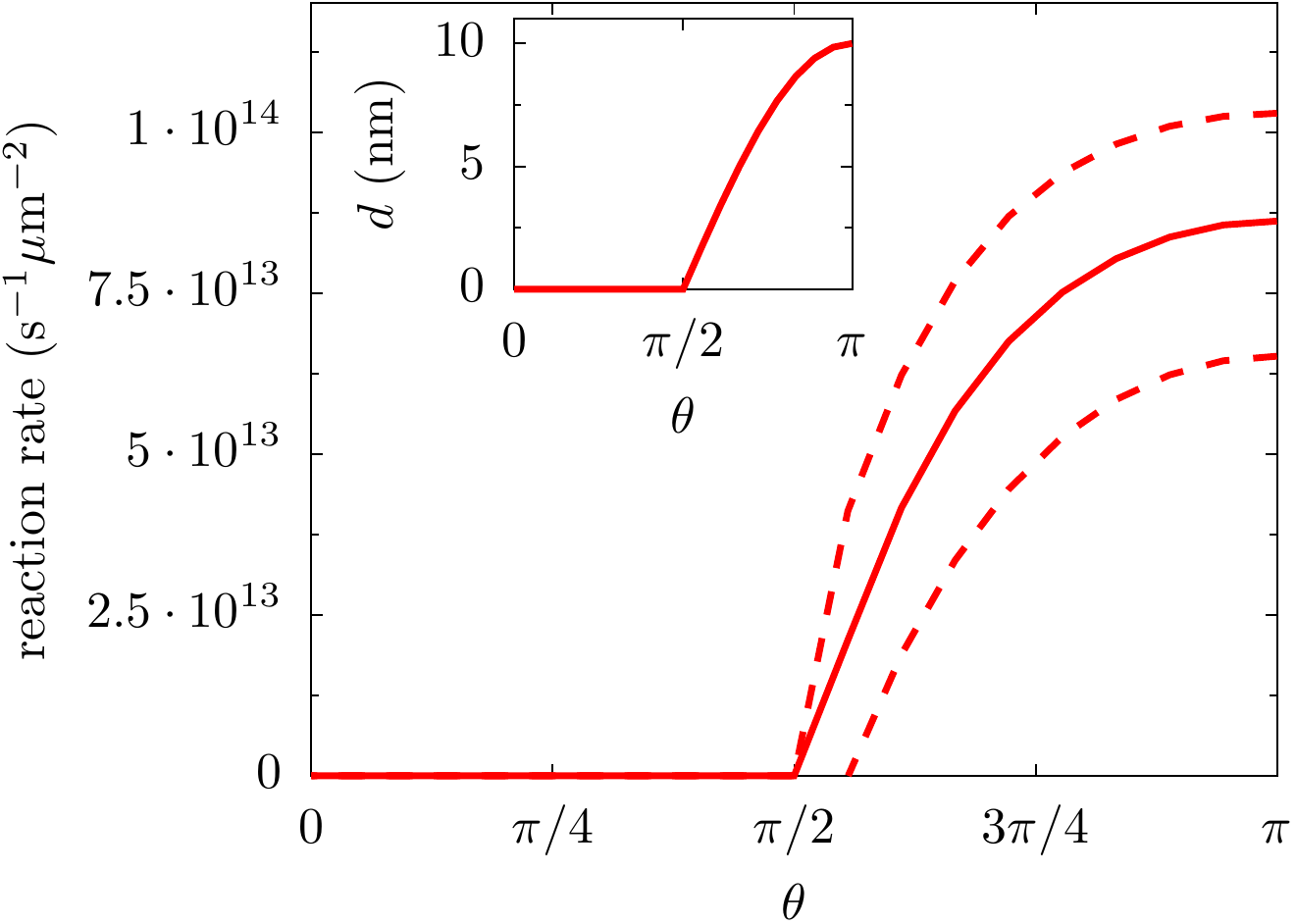}
\caption{Predicted variation in catalytic activity across the Janus sphere surface based on an exponential fit to reactivity measurements reported in \cite{Howse2007} and geometrically determined variation in Platinum coating thickness (inset).  The dashed lines represent  $\pm$ 1 standard deviation of the fit used to interpolate the reactivity data in \cite{Howse2007}.}
\label{Pt_thickness}
\end{center}
\end{figure}

{It is also interesting to compare this simplified slip-velocity profile, which is consistent with the results obtained from the above fitting procedure to the full experimental flow field, with estimated pole to equator changes in Pt surface reactivity as a function of $\theta$.  These reactivity variations can be related to the Pt coating thickness ($d$) variations \cite{Howse2007} shown in Figure \ref{Pt_thickness} inset ($d$ can be calculated by a simple geometric model: $d=d_\text{max} \cos(\pi-\theta)$ where $d_\text{max}$ is 10 nm in our experiment, and $\pi-\theta$ gives the angle between the incident Pt ``rays'' during metal evaporation and the surface normal vector for the sphere).  Using an exponential fit (reactivity = $A + B \text{e}^{cd}$; $A=1.12 \cdot10^{14} \pm 8\cdot 10^{12} \text{s}^{-1} \mu\text{m}^{-2}$, $B=-1.18 \cdot 10^{14}  \pm 9.67\cdot 10^{12} \text{s}^{-1} \mu \text{m}^{-2}$, $C=-0.15236\pm0.0329$) to the previously published experimental data for variations in surface reactivity as a function of Pt thickness from \cite{Howse2007}, it is consequently possible to arrive at predictions for the reactivity variation for the catalytically active hemisphere, (Fig. \ref{Pt_thickness}).}

\begin{figure}
\begin{center}
\includegraphics[width=0.8\columnwidth]{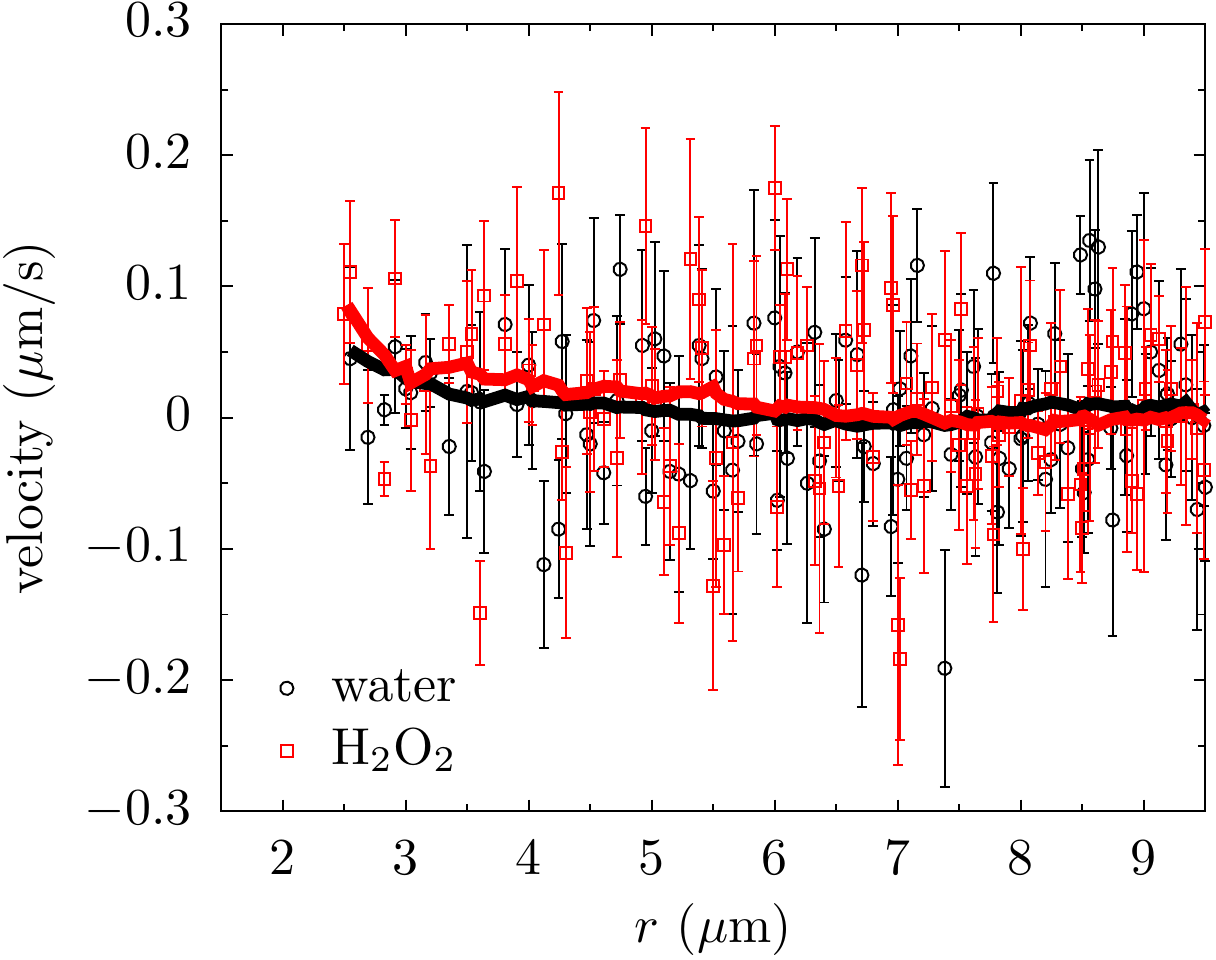}
\caption{Comparison of the radially averaged velocities of tracer particles as a function of radial distance from the centre of a stuck symmetrically Pt coated colloid in water, and in 10 \% $\text{H}_2\text{O}_2$.  The solid lines are adjacent averages with a window of 5 points to aid the eye. 
}
\label{control}
\end{center}
\end{figure}

We show the experimentally measured velocity fields for tracer particle motion in Figs. \ref{streamlines}(c) and  \ref{streamlines}(d) (see \cite{SI} for zoom-out versions). According to Faxen law, the tracers' velocity will reflect both the fluid flow, and any other forces acting on the particles.  The potential for the chemical gradient generated by the catalytic reaction to exert a phoretic force on the tracer clearly exists in the system under investigation.  Consequently, to assess the magnitude of any phoretic contribution to tracer motion, a control experiment was performed to observe the motion of tracer particles in the vicinity of a stuck, symmetrically catalytically active colloid, fully covered in Pt.  In this case, the symmetry of the catalytic reaction precludes the generation of flow fields, and the chemical gradient effect can be measured in isolation.  Figure \ref{control} displays a graph depicting the mean radially averaged velocity of the tracer particles as a function of radial distance from the symmetrical gradient only Pt colloid, both in water and in the presence of 10 \% w/v hydrogen peroxide fuel.  It is clear that within error the tracer velocities are unchanged by the presence of the chemical gradient instigated by the addition of the hydrogen peroxide.  Based on this evidence, the subsequent discussions considers the tracers' velocity field to be equivalent to the fluid flow field.

\begin{figure}
\begin{center}
\includegraphics[width=\columnwidth]{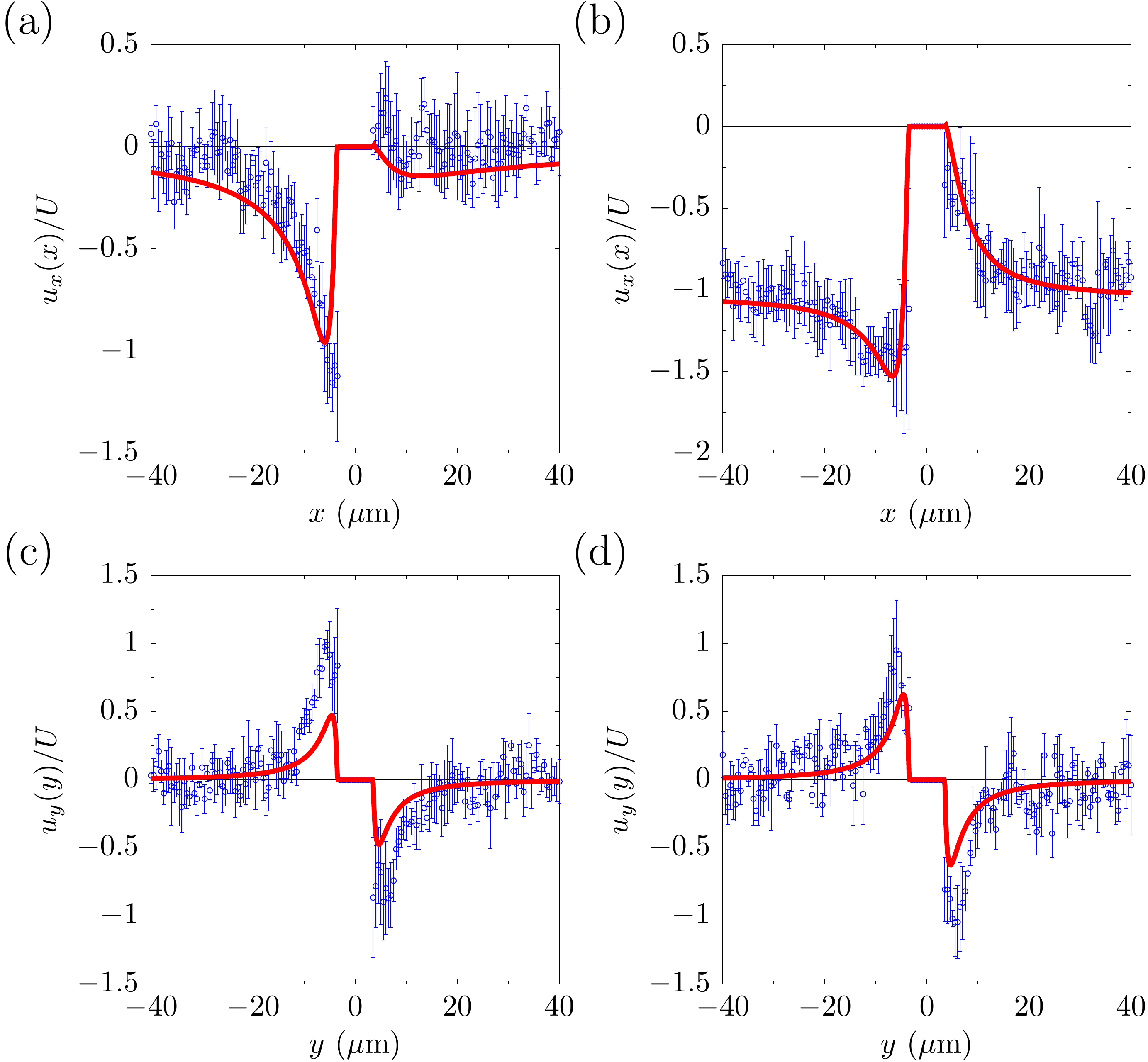}
\caption{Top: $x$-component of the velocity measured at $y=0$ and as a function of the coordinate $x$ for the two situations where the swimmer is stuck (a) and freely moving (b). Bottom: $y$-component of the velocity measured at $x=0$ and as a function of the coordinate $y$ for the two situations where the swimmer is stuck (c) and freely moving (d). For all plots, the blue symbols and error bars are obtained experimentally after averaging the fields measured in a stripe of width $\unit[2]{\mu\mathrm{m}}$ around $y=0$ (top) and $x=0$ (bottom). The red line is a fit of the experimental data using the expressions of the velocity given in the Supplementary Information \cite{SI}, with $K=v_0/U$ as a free parameter. The expansions in Legendre polynomials are truncated at order $N=25$. The values of the fit parameters are $K_\text{stuck} = 11.6\pm0.7$ and $K_\text{moving}=15.2\pm0.7$.}
\label{along_axes}
\end{center}
\end{figure}

It is apparent that the stuck particle [Fig. \ref{streamlines}(c)] acts as a catalytic pump.  Streamlines and velocity magnitudes indicate that the fluid is drawn in towards the equatorial region of the static Pt-PS particle and then pushed away from the Pt catalyst coated hemisphere. The pumping direction (away from Pt cap) is consistent with the direction of travel observed here, and reported previously \cite{Ebbens2011}. In regions of low velocity, such as in the fluid surrounding the inactive side of the Pt-PS particle, Brownian translations of the tracer particles influence the streamlines producing stochastic detail, whereas the catalytically induced velocity dominates elsewhere producing smooth streamlines. The flow field around the moving colloid is consistent with the stuck colloid \cite{SI}, [Fig. \ref{streamlines}(d)] exhibiting a tendency for fluid to be drawn in towards the equator of the particle.  However, due to the overall translation of the fluid relative to the frame of reference, the movement towards the particle is less marked.  The background relative fluid flow dominates the tracer particle Brownian motion, leading to smooth streamlines. The flow field retains the expected symmetry about the $y=0$ line, which is a good indication that the transformation of tracer particle motion required to arrive at a fixed frame of reference, and averaging over multiple moving Pt-PS particles have been effective in capturing the essential details of the flow.  The overall velocity magnitude variations are less striking compared to the stuck particle.  However, a small area of reduced velocity at the pole of the Pt cap, and a larger area of low velocity around the PS cap are qualitatively apparent, together with an overall increased average velocity on the Pt side of the particle.  Comparison with the theoretical predictions finds a good qualitative agreement between the structures of the flow field around the swimmer. In order to check quantitative agreement, and to observe more thoroughly the decay of the flow field far away from the surface of the colloid, we study different projections of the velocity field, namely $u_x$ as a function of $x$ around $y=0$ and $u_y$ as a function of $y$ around $x=0$ for the two situations (Fig. \ref{along_axes}). To improve the statistics of the measurements, we average the velocity fields over stripes of width 2 $\mu$m around $y=0$ and $x=0$. These results are fitted with the components of the flow fields computed analytically with $K=v_0/U$ [Eq. \eqref{slip}] as a free parameter using a least-square method, for both the moving and stuck cases. We obtain $K_\text{stuck} = 11.6\pm0.7$ and $K_\text{moving}=15.2\pm0.7$. These estimates are comparable to the theoretical prediction of $K=14.6$.

\textit{Influence of the solid wall.---} To the best of our knowledge, the generic solution for the flow field around a Janus colloid with a prescribed slip velocity $v(\theta)$ and near a wall with no-slip has not been established analytically. Alternatively, Spagnolie and Lauga \cite{Spagnolie2012} proposed to represent the flow field at position $\rr$ around a squirmer as the superposition of the flow fields created by elementary singularities located at the centre of the colloid $\rr_0$ (respectively Stokeslet dipole, source dipole and Stokeslet quadrupole at order $\mathcal{O}(|\rr-\rr_0|^4)$ in the distance to centre of the the colloid) whose relative strengths depend on the Legendre coefficients of the slip velocity $B_n$ \cite{Ishimoto2013}. More precisely, in units of $U$ and in the case of a freely swimming colloid, they establish
\begin{eqnarray}
\uu/U &=& -\ee +\alpha (\boldsymbol{G_D}+\boldsymbol{G_D}^*) + \beta a (\boldsymbol{D}+\boldsymbol{D}^*) \nonumber\\
&&+ \gamma a(\boldsymbol{G_Q}+\boldsymbol{G_Q}^*) + \boldsymbol{O}(|\rr-\rr_0|^4)
\label{sing_Lauga}
\end{eqnarray}
where the quantities $\boldsymbol{G_D}$, $\boldsymbol{D}$ and $\boldsymbol{G_Q}$  represent respectively the Stokeslet dipole, source dipole and Stokeslet quadrupole (evaluated at $\rr-\rr_0$), and the starred quantities their mirror-image (evaluated at $\rr-\rr_0^*$) where $\rr_0^*$ is the symmetric of the centre of the colloid with respect to the wall. All of these functions depend on the orientation $\ee$ of the colloid. $\alpha$, $\beta$ and $\gamma$ are dimensionless coefficients that depend on $B_1$, $B_2$ and $B_3$ \cite{Ishimoto2013,SI}. 

\begin{figure}
\begin{center}
\includegraphics[width=\columnwidth]{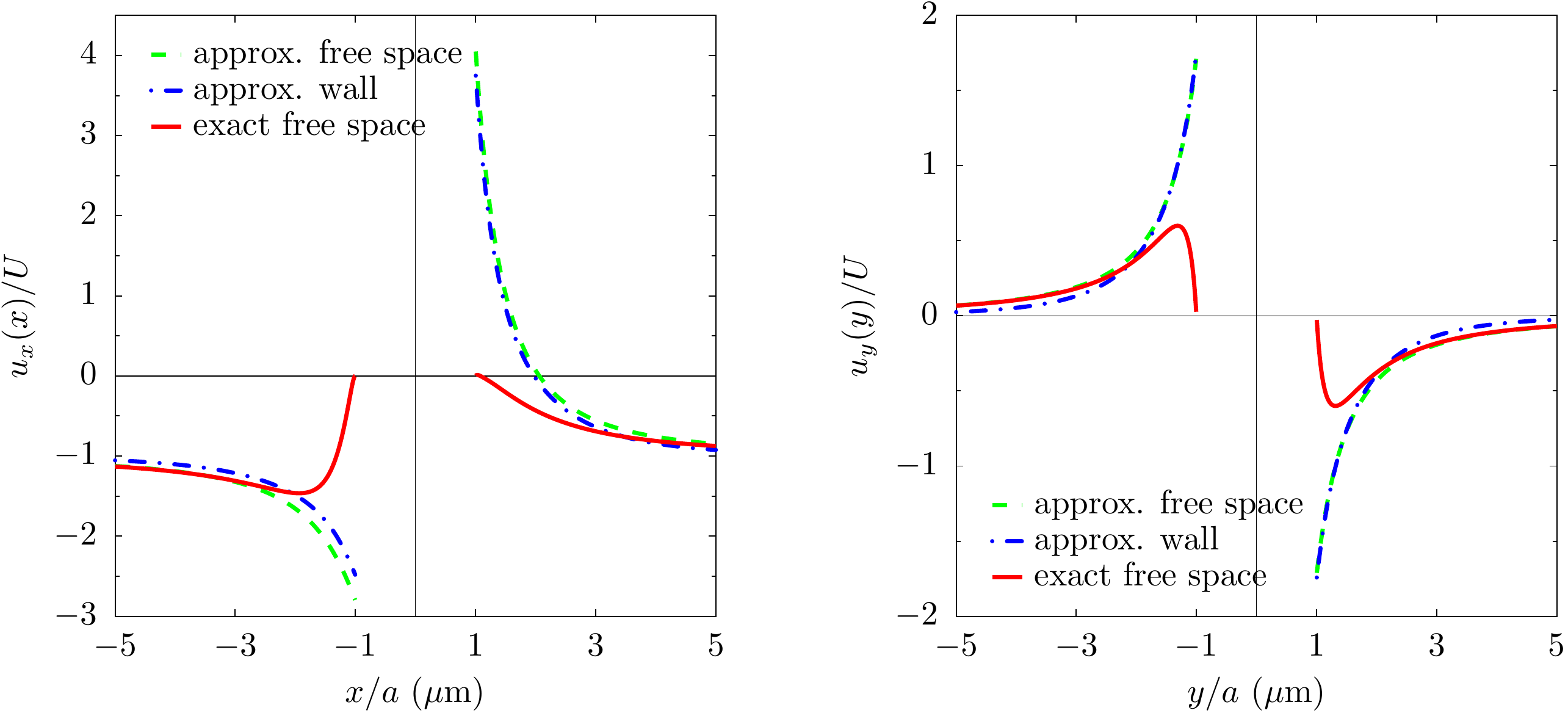}
\caption{Components of the flow field [left: $u_x(x,y=0)$; right: $u_y(x=0,y)$] in the equatorial plane of the Janus sphere, calculated using the expression obtained by the singularities expansion in free space \cite{SI}, the singularities expansion and their images that account for the presence of a no-slip wall [Eq. \eqref{sing_Lauga}] and the exact solution in free space, for the boundary condition at the surface of the colloid $v(\theta)$ given in Eq. \eqref{slip}.}
\label{fig_wall}
\end{center}
\end{figure}

As opposed to the full solution of the Stokes equation in free-space, which is valid at any point \cite{Lighthill1952, Blake1971a}, this solution is a far-field expansion, so we expect it to be more accurate further away from the colloid. Using the simplified shape of $v(\theta)$ [Eq. \eqref{slip}], we can compare these two calculated flow field. As examples, we consider two projections of the flow field along the main axes of the colloid ($u_x(x,y=0)$ and $u_y(x=0,y)$) (Fig. \ref{fig_wall}). As expected, the two solutions coincide far away from the surface of the colloid, and we observe that the singularities expansion of the flow field fails to describe faithfully the behaviour of $\uu$ in the close vicinity of the colloid. The difference between the two fields far away from the tracer is within the noise of the data recorded. This comparison justifies the choice of the free-space solution. A far-field expansion that would account for the presence of the wall would only be useful to probe very fine differences, that cannot be reached within experimental precision.

\paragraph*{Discussion.---}The very good quantitative agreement between the components of the velocity measured analytically and experimentally shows that the simple shape of the slip velocity [Eq. \eqref{slip} and Fig. \ref{streamlines}(b)] correctly captures the decay of the velocity field from the surface all the way to regions far from the Pt-PS particle, which is strongly in favor of the mechanism proposed in \cite{Ebbens2014,Das2015}. The flow fields around a stuck or a moving swimmer for other candidate slip velocity profiles do not show a similarly good agreement in comparison \cite{SI}. A slip velocity that is fore-aft symmetric with a peak at the equator and that vanishes at the poles, which corresponds to a source dipole, does not yield the right structure for the flow field (in particular because the projection of $\uu$ along $y$ vanishes at $x=z=0$). A profile of slip velocity that would be uniform on the Pt hemisphere and zero on the PS hemisphere will yield velocity profiles that are closer to the ones observed experimentally (even though there is no underlying theoretical mechanism to support this profile). 

{Our observations are consistent with a velocity profile that behaves as a pusher in the far-field limit \cite{SI}, which which can be a starting point for a description of the hydrodynamic interactions between many such self-phoretic active colloids \cite{Julicher2009,Tu2013,Yang2014,Kreissl2016}. Indeed, using the usual definition of the squirmer parameter $b=B_2/|B_1|$, we find from the data fit (inset of Fig. \ref{streamlines}(b)) $b\simeq-2.45<0$, which corresponds to a pusher.} However, our experiments have provided a more complete picture with regards to the near-field properties of the hydrodynamic interactions than a simplistic squirmer of pusher type, which can be used to build a more faithful representation of the hydrodynamic interactions. Therefore, our results will pave the way for a more comprehensive description of the self-organization of such active colloids through hydrodynamic and phoretic interactions \cite{{Zottl2016a},{Elgeti2015}}.

In conclusion, we have examined the hydrodynamic flow fields around a self-propelled Pt-PS catalytic active colloid and found flow fields that are strongly asymmetric across the surface of the particle, with larger magnitude on the Pt-coated hemisphere. The experimental measurements show good agreement with the explicit solution of the Stokes equation with a simple slip velocity as a boundary condition, which is consistent with the conjectured current loops. Our results can be used to construct a comprehensive theoretical description of a suspension of such active colloids.\\

\begin{acknowledgements}
Both AIC and SJE thank the EPSRC for supporting this work via the Career Acceleration Fellowship (EP/J002402/1) granted to SJE. PI and RG would like to thank the hospitality of the Max-Planck-Institute for the Physics of Complex Systems, where a part of this work was performed. We acknowledge support from the COST Action MP1305 Flowing Matter.
\end{acknowledgements}


\onecolumngrid

\pagebreak

\beginsupplement

\begin{center}

\textbf{Experimental Observation of Flow Fields Around Active Janus Spheres}

$\ $

\textit{\textbf{Supplementary Information}}

$\ $

Andrew I. Campbell,$^1$ Stephen J. Ebbens,$^{1,*}$ Pierre Illien,$^2$ and Ramin Golestanian$^{3, 4, \dagger}$

$\ $

$^1$\emph{Department of Chemical and Biological Engineering, University of Sheffield, Mappin Street, Sheffield S1 3JD, UK}

 $^2$\emph{Sorbonne Universit\'e, CNRS, Laboratoire PHENIX, 4 place Jussieu, 75005 Paris, France}

$^3$\emph{Max Planck Institute for Dynamics and Self-Organization (MPIDS), 37077 G\"ottingen, Germany}

$^4$\emph{Rudolf Peierls Centre for Theoretical Physics, University of Oxford, Oxford OX1 3PU, United Kingdom}

\end{center}


\section{Solution of the Stokes problem}

\subsection{Stuck swimmer}
\label{sec_stuck}

The total flow field around a stuck Janus swimmer reads
\begin{equation}
\label{ }
\uu_\text{s} = \uu^{\text{ph}} + \uu^{\text{mono}},
\end{equation}
where
\begin{itemize}
  \item $\uu^{\text{ph}}$ is the phoretic contribution (whether electrophoretic or diffusiophoretic),
  \item $\uu^{\text{mono}}$ is the contribution from the force monopole that maintains the colloid steady in the laboratory reference of frame.
\end{itemize}

$\uu^{\text{ph}}$ is the solution of the Stokes problem [Eq. (1) from the main text] 
with the  boundary condition
\begin{equation}
u^\text{ph}_\theta |_{r=a}  =v(\theta)=  \sum_{n = 1}^\infty B_n V_n(\cos\theta),\label{Bn}
\end{equation}
where
\begin{equation}
\label{ }
V_n(\cos\theta) = \frac{2}{n(n+1)}\sin \theta\,  P'_n(\cos \theta),
\end{equation}
{where $P_n$ are the Legendre polynomials.} The contribution from the force monopole  $\ff = -f \ee$ ($f>0$) is
\begin{equation}
\label{ }
\uu^{\text{mono}} = \frac{1}{2} \frac{f}{6\pi\eta a} \cos(\theta) \left[ \left(\frac{a}{r}\right)^3 -3 \left(\frac{a}{r}\right) \right] \ee_r
+\frac{1}{4} \frac{f}{6\pi\eta a} \sin(\theta) \left[ \left(\frac{a}{r}\right)^3 +3 \left(\frac{a}{r}\right) \right] \ee_\theta,
\end{equation}
or, in dyadic notations,
\begin{equation}
\label{umono}
\uu^{\text{mono}} = \frac{1}{2} \frac{f}{6\pi\eta a}  \left[ \left(\frac{a}{r}\right)^3 -3 \left(\frac{a}{r}\right) \right] \frac{\ee \cdot \rr}{r}  \frac{\rr}{r}
+\frac{1}{4} \frac{f}{6\pi\eta a}  \left[ \left(\frac{a}{r}\right)^3 +3 \left(\frac{a}{r}\right) \right] \left(\frac{\ee \cdot \rr}{r}  \frac{\rr}{r} - \ee \right).
\end{equation}
Using the known expression of $\uu^\text{ph}$ \cite{Lighthill1952,Blake1971a}, the total flow field then writes
\begin{eqnarray}
\uu_\text{s}  = && - \frac{1}{3} \frac{a^3}{r^3} B_1  \ee +B_1 \frac{a^3}{r^3} \frac{\ee\cdot \rr}{r} \frac{\rr}{r} + \sum_{n=2}^\infty \left( \frac{a^{n+2}}{r^{n+2}} - \frac{a^n}{r^n} \right) B_n P_n\left( \frac{\ee\cdot\rr}{r} \right) \frac{\rr}{r} \nonumber \\
&&+\sum_{n=2}^\infty \left(  \frac{n}{2} \frac{a^{n+2}}{r^{n+2}}- \left(\frac{n}{2}-1\right)\frac{a^n}{r^n}\right) B_n W_n\left( \frac{\ee\cdot\rr}{r}\right) \left( \frac{\ee\cdot\rr}{r} \frac{\rr}{r} -\ee\right) \nonumber \\
&&+\frac{1}{2} \frac{f}{6\pi\eta a}  \left[ \left(\frac{a}{r}\right)^3 -3 \left(\frac{a}{r}\right) \right] \frac{\ee \cdot \rr}{r}  \frac{\rr}{r}
+\frac{1}{4} \frac{f}{6\pi\eta a}  \left[ \left(\frac{a}{r}\right)^3 +3 \left(\frac{a}{r}\right) \right] \left(\frac{\ee \cdot \rr}{r}  \frac{\rr}{r} - \ee \right),
\end{eqnarray}
with
\begin{equation}
\label{ }
W_n(\cos\theta) \equiv \frac{V_n(\cos\theta)}{\sin\theta}  = \frac{2}{n(n+1)}  P'_n(\cos \theta).
\end{equation}
The flow field at the surface of the colloid is
\begin{equation}
\label{ }
\uu|_{\rr= a \ee_r} = - \frac{f}{6\pi \eta a}\ee + \frac{2}{3}B_1 \ee + \sum_{n \geq 2} B_n V_n(\cos \theta) \ee_\theta.
\end{equation}
Imposing $f$ such that the swimmer does not move, we find $f=4\pi\eta a B_1$, and
\begin{empheq}[box=\fbox]{align}
\uu_\text{s}(r,\theta) =& \frac{1}{2} B_1 \left\{ -\frac{a}{r} \left[  \ee + \frac{\ee \cdot \rr}{r} \frac{\rr}{r} \right]  + \left(\frac{a}{r} \right)^3 \left[  3 \frac{\ee \cdot \rr}{r} \frac{\rr}{r} {-\ee} \right]\right\} + \sum_{n=2}^\infty \left( \frac{a^{n+2}}{r^{n+2}} - \frac{a^n}{r^n} \right) B_n P_n\left( \frac{\ee\cdot\rr}{r} \right) \frac{\rr}{r} \nonumber\\
&+\sum_{n=2}^\infty \left(  \frac{n}{2} \frac{a^{n+2}}{r^{n+2}}- \left(\frac{n}{2}-1\right)\frac{a^n}{r^n}\right) B_nW_n\left( \frac{\ee\cdot\rr}{r}\right) \left( \frac{\ee\cdot\rr}{r} \frac{\rr}{r} -\ee\right)
\label{field_eq_stuck}
\end{empheq}

\begin{figure}[t]
{\raisebox{60mm}{\large{(a)}}{\includegraphics[width=7.8cm]{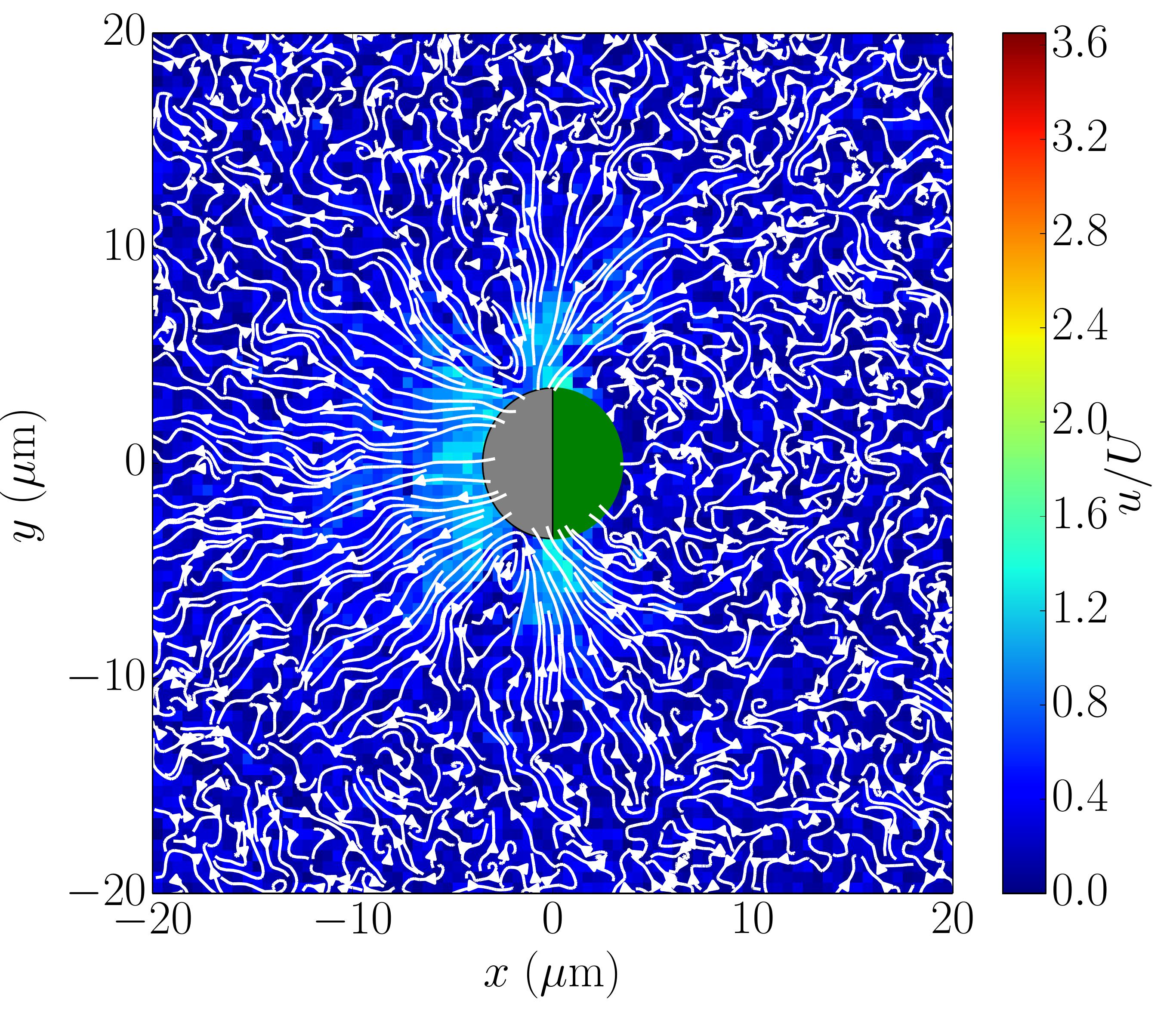}}}
{\raisebox{60mm}{\large{(b)}}{\includegraphics[width=7.8cm]{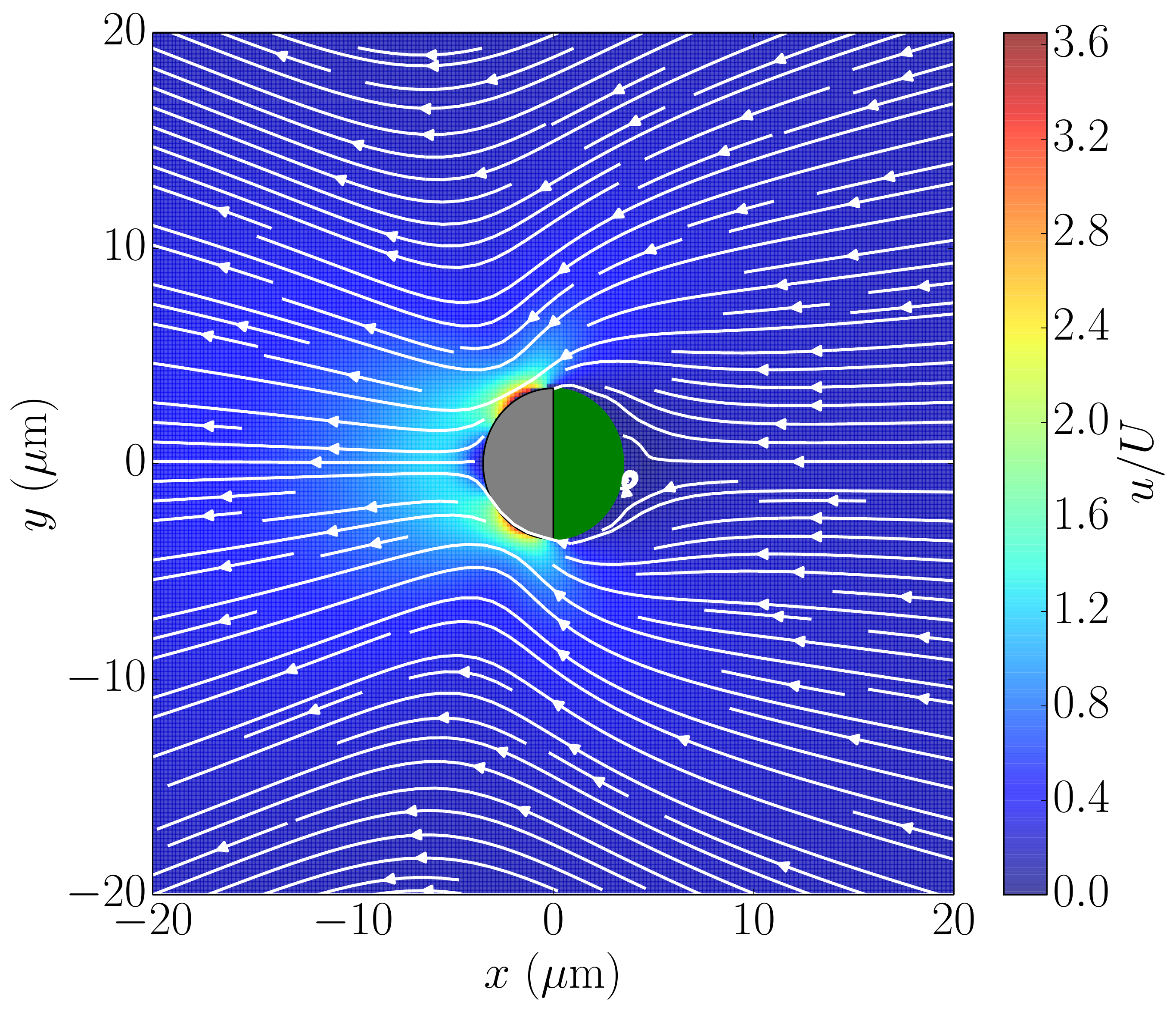}}}\\
{\raisebox{60mm}{\large{(c)}}{\includegraphics[width=7.8cm]{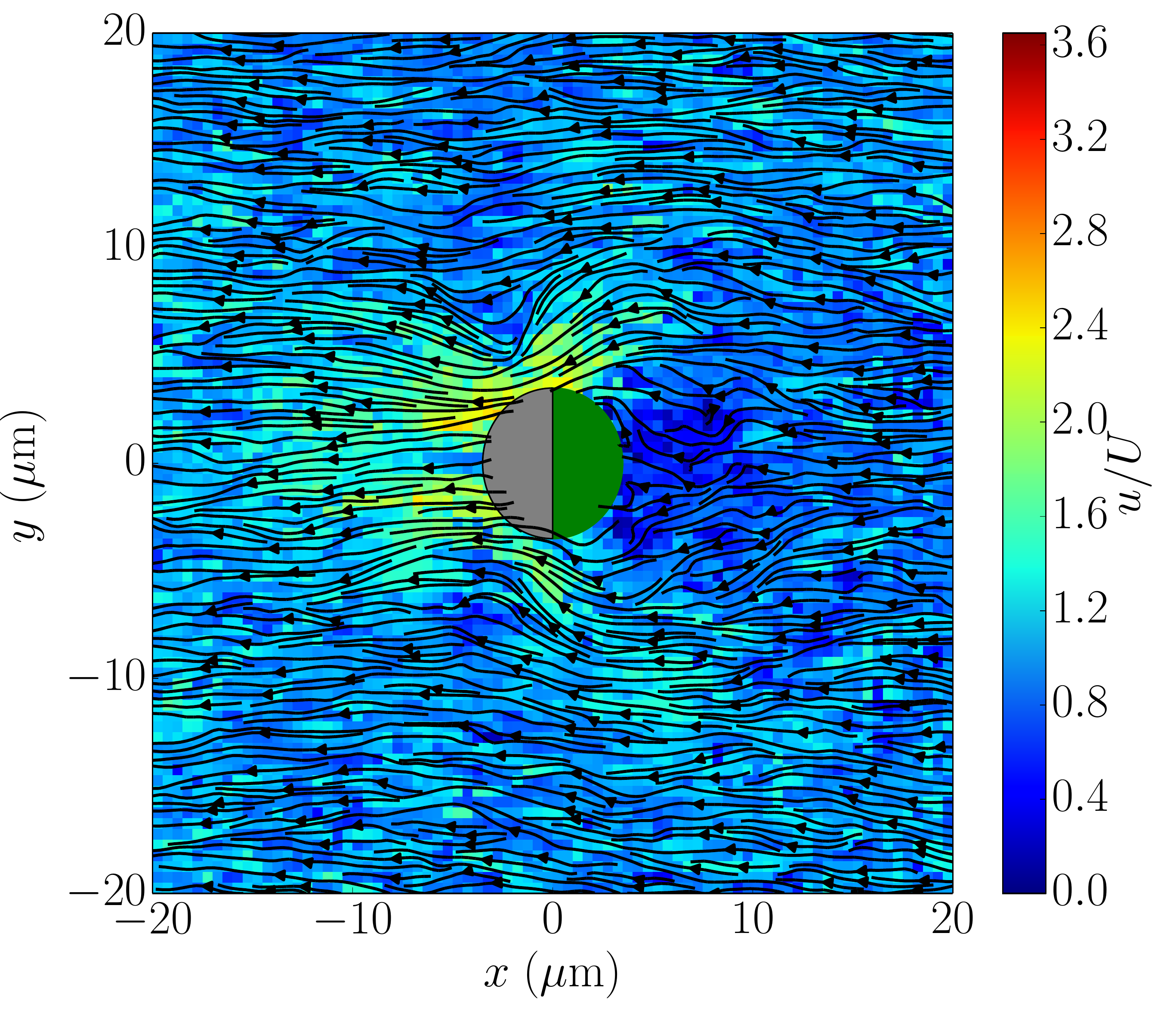}}}
{\raisebox{60mm}{\large{(d)}}{\includegraphics[width=7.8cm]{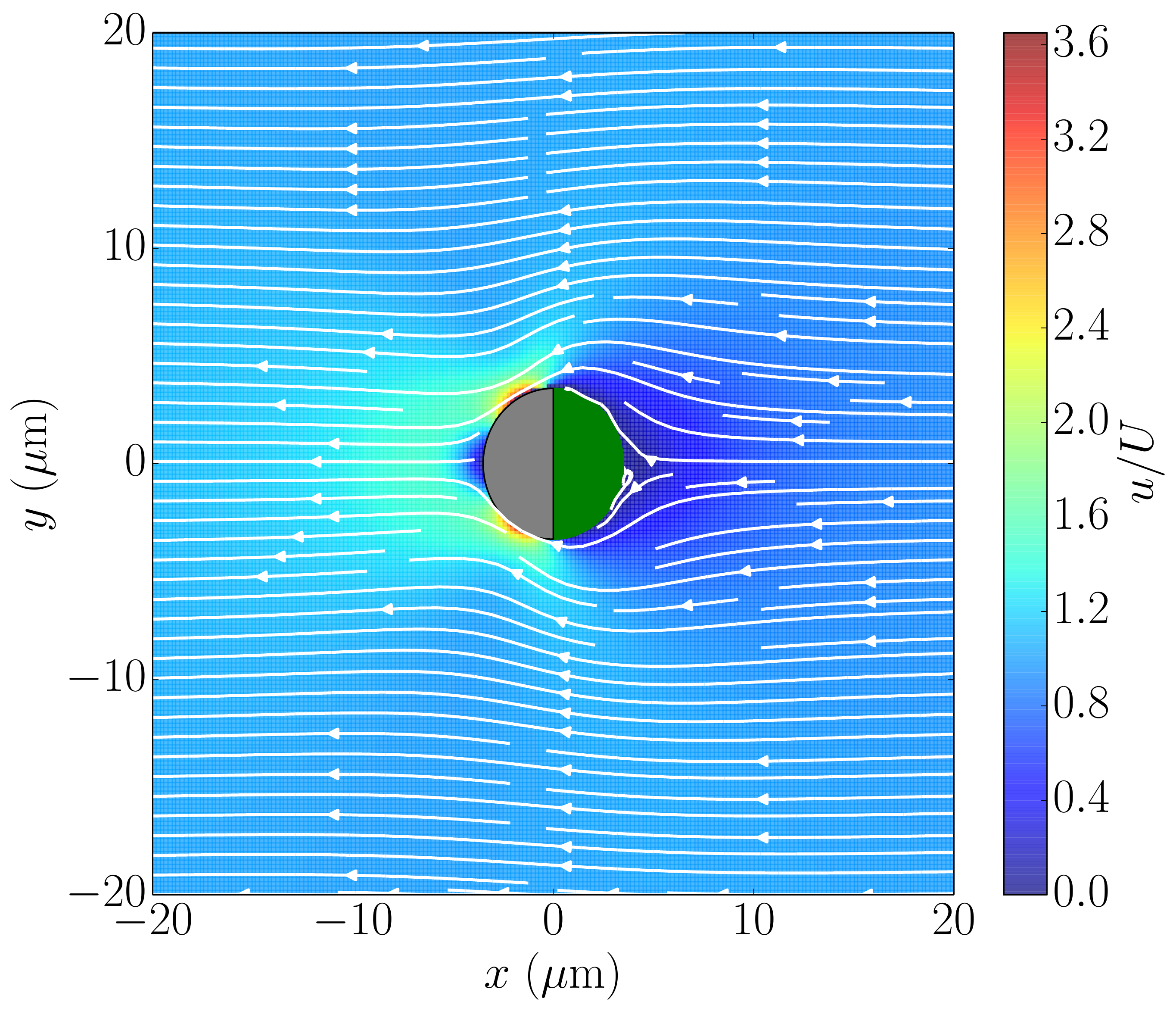}}}
\caption{Zoomed-out version of Fig. 1. (a)-(d) Streamlines around the Janus particles obtained experimentally (left) and analytically (right), in the two situations where the particle is stuck (top) and freely moving (bottom). The background colors represent the magnitude of the velocity $u=|\uu|$ rescaled by the swimming velocity of the Janus particle when it is freely moving.}
\label{zoomedout}
\end{figure}

\subsection{Moving swimmer: velocity field in the co-moving frame}

Ignoring the contribution coming from the force monopole and solving the Stokes problem with the boundary condition
\begin{equation}
\lim_{r\to\infty} \uu  =  -U \ee,
\end{equation}
where $U=\frac{2}{3} B_1$ is the swimming velocity of the colloid, we find
\begin{empheq}[box=\fbox]{align}
\uu_\text{m}(r,\theta)   = &-\frac{2}{3}B_1 \ee - \frac{1}{3} \frac{a^3}{r^3} B_1 \ee + B_1 \frac{a^3}{r^3} \frac{\ee\cdot \rr}{r} \frac{\rr}{r} + \sum_{n=2}^\infty \left( \frac{a^{n+2}}{r^{n+2}} - \frac{a^n}{r^n} \right) B_n P_n\left( \frac{\ee\cdot\rr}{r} \right) \frac{\rr}{r} \nonumber \\
&+\sum_{n=2}^\infty \left(  \frac{n}{2} \frac{a^{n+2}}{r^{n+2}}- \left(\frac{n}{2}-1\right)\frac{a^n}{r^n}\right) B_n W_n\left( \frac{\ee\cdot\rr}{r}\right) \left( \frac{\ee\cdot\rr}{r} \frac{\rr}{r} -\ee\right),
\label{field_eq_moving}
\end{empheq}

\section{Enlarged view of the streamlines around the Janus particles}

In Fig. S1, 
we show a zoomed-out representation of the plots presented in Fig. 1 
for $x$ and $y$ varying from $-20$ $\mu$m to  $20$ $\mu$m.

\section{Fitting procedure}

We start from the expression of the flow field in terms of the Legendre coefficients associated to the slip velocity [Eqs. \eqref{field_eq_stuck} for the case where the colloid is stuck and \eqref{field_eq_moving} for the case where it is freely moving]. From the experimental data, we have actually four independent sets of data: $\uu_{\text{s}} \cdot \ee_r(r,\theta)$, $\uu_{\text{s}} \cdot \ee_\theta(r,\theta)$, $\uu_{\text{m}} \cdot \ee_r(r,\theta)$ and $\uu_{\text{m}} \cdot \ee_\theta(r,\theta)$. Projecting Eqs. \eqref{field_eq_stuck} and \eqref{field_eq_moving} onto the unit vectors $\ee_r$ and $\ee_{\theta}$, we find the corresponding fit functions (we introduce for simplicity $X\equiv a/r$):
\begin{eqnarray}
\uu_{\text{s}} \cdot \ee_r(X,\theta) & = & \sum_{n=1}^\infty B_n (X^{n+2}-X^n)P_n(\cos\theta) \\
\uu_{\text{s}} \cdot \ee_\theta(X,\theta) & = & \sin\theta  \sum_{n=1}^\infty \left[ \frac{n}{2}X^{n+2}-\left(\frac{n}{2}-1 \right) X^n  \right] \frac{2B_n}{n(n+1)}P'_n(\cos\theta)\\
\uu_{\text{m}} \cdot \ee_r(X,\theta) & = & \frac{2}{3}(X^3-1)B_1\cos\theta+\sum_{n=2}^\infty B_n (X^{n+2}-X^n)P_n(\cos\theta) \\
\uu_{\text{m}} \cdot \ee_\theta(X,\theta) & = & \frac{1}{3}B_1(2+X^3)\sin\theta+ \sin\theta  \sum_{n=1}^\infty \left[ \frac{n}{2}X^{n+2}-\left(\frac{n}{2}-1 \right) X^n  \right] \frac{2B_n}{n(n+1)}P'_n(\cos\theta)
\end{eqnarray}

Each of these functions is used to fit the experimental data (for $(x,y)\in [-40~\mu\text{m},40~\mu\text{m}]^2$, with the constraint $r\geq 3.5~\mu\text{m}$). The sums are truncated at order 5. Therefore there are 5 fit parameters $B_1,\dots,B_5$. The values obtained from the fits (as well as the averages and standard deviations) are shown on the inset of Fig. 1(a). We can then rebuild the slip velocity at the surface of the colloid, defined as
\begin{equation}
v(\theta) =  \sum_{n = 1}^N B_n V_n(\cos\theta),\label{Bn}
\end{equation}
where
\begin{equation}
\label{ }
V_n(\cos\theta) = \frac{2}{n(n+1)}\sin \theta\,  P'_n(\cos \theta).
\end{equation}

The reconstructed $v(\theta)$ is shown on the main plot of Fig. 1(a).

\section{Influence of the solid wall}

\begin{figure}[h]
\begin{center}
\includegraphics[width=12cm]{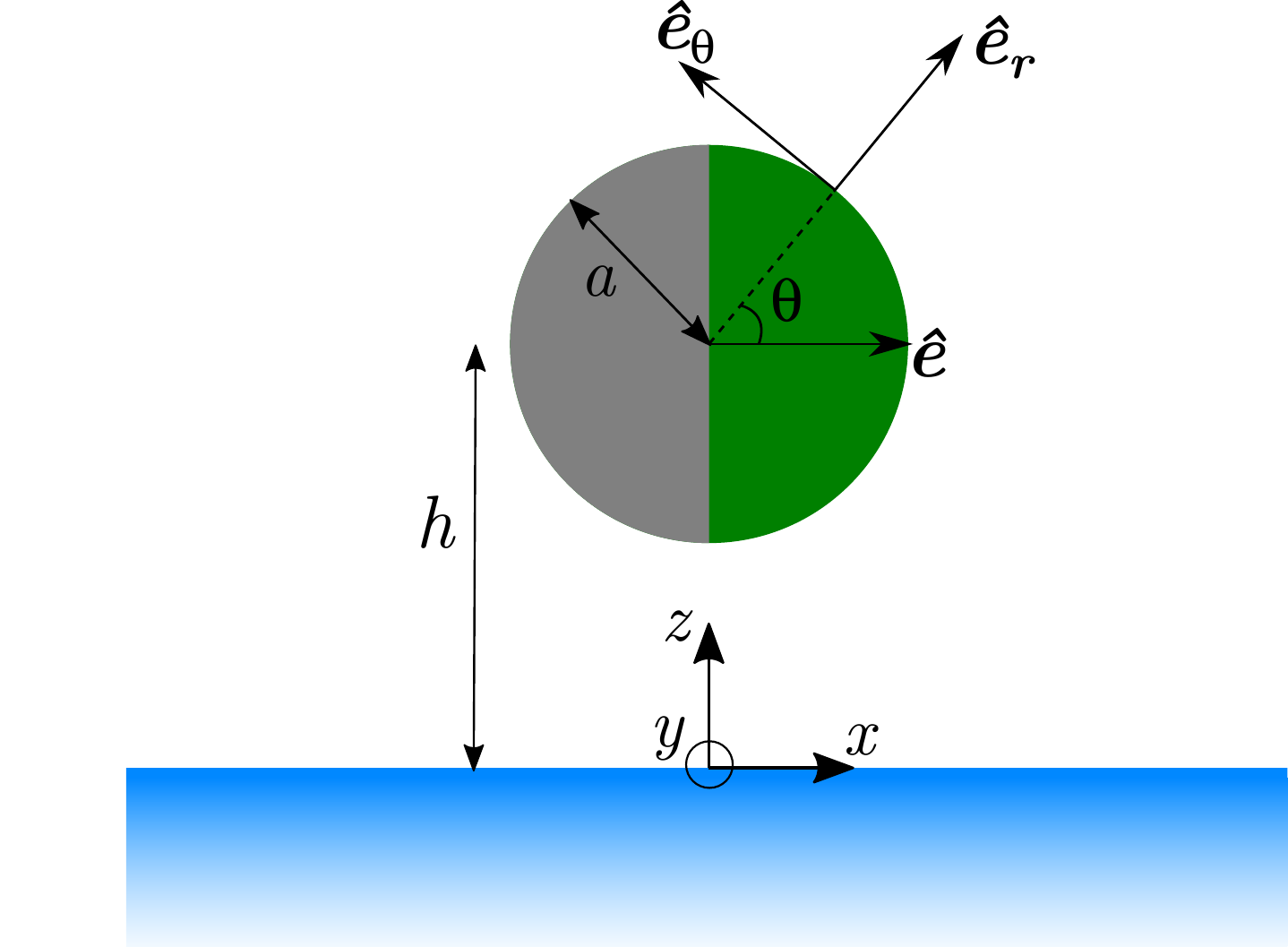}
\caption{Notations for the singularities calculation}
\label{notations}
\end{center}
\end{figure}

In this section, the origin of coordinates is a point on the solid wall (see Fig. \ref{notations}). Following Spagnolie and Lauga \cite{Spagnolie2012}, the flow field created by a squirmer whose centre is at position $\rr_0 = (0,0,h)$ and with zero rotational velocity can be decomposed into a sum of singularities which correspond to a far field expansion (\cite{Spagnolie2012} Eq. (2.17)):
\begin{equation}
\boxed{
\widetilde{\uu}_\text{sing,free}(\rr) = -\ee+ \alpha \boldsymbol{G_D} (\rr-\rr_0,\ee,\ee)+ \beta a \boldsymbol{D}(\rr-\rr_0,\ee)+\gamma a \boldsymbol{G_Q}(\rr-\rr_0,\ee,\ee)+\mathcal{O}(|\rr-\rr_0|^4) 
}
\label{sing_free}
\end{equation}
(note that this is dimensionless i.e. the velocity is measured in units of $2B_1/3$ and Lauga's result was modified to be in the frame of reference attached to the colloid). Here $\alpha,\beta$ and $\gamma$ are dimensionless coefficients that characterize the strength of each singularity, and one defines:
\begin{itemize}
  \item the Stokeslet dipole $\boldsymbol{G_D}$, see \cite{Spagnolie2012} Eq. (A2),
  \item the source dipole $\boldsymbol{D}$, see \cite{Spagnolie2012} Eq. (A8),
  \item the Stokeslet quadrupole $\boldsymbol{G_Q}$ \cite{Spagnolie2012} Eq. (A3).
\end{itemize}

Ishimoto and Gaffney \cite{Ishimoto2013} give the relation between $\alpha,\beta$ and $\gamma$  and the Legendre coefficients of the slip velocity at the surface of the colloid (see  \cite{Ishimoto2013} Eq. (7)):
\begin{eqnarray}
\alpha & = & - \frac{3}{4} \frac{B_2}{B_1}\\
\beta & = & \frac{1}{2}- \frac{1}{8} \frac{B_3}{B_2}\ \\
\gamma & =& - \frac{5}{16} \frac{B_3}{B_2}\
\end{eqnarray}
They are indeed dimensionless since each of the $B_n$ has the dimension of a velocity. We note that because it relies on a singularity expansion, the expression of $\widetilde{\uu}$ in Eq. \eqref{sing_free} only involves the first three Legendre coefficients of the slip velocity $B_1$, $B_2$ and $B_3$. Any higher-order contribution to the slip velocity won't have any effect on the computed $\widetilde{\uu}$ at this level of expansion.\\

Applying the method of images, the flow field in the presence of the wall is obtained from $\widetilde{\uu}$ in Eq. \eqref{sing_free} by adding to each of the singularities its image:
\begin{empheq}[box=\fbox]{align}
\widetilde{\uu}_\text{sing,wall}(\rr)  = -\ee&+ \alpha \boldsymbol{G_D} (\rr-\rr_0,\ee,\ee)+ \beta a \boldsymbol{D}(\rr-\rr_0,\ee)+\gamma a \boldsymbol{G_Q}(\rr-\rr_0,\ee,\ee)\nonumber \\
&+ \alpha \boldsymbol{G_D}^* (\rr-\rr_0^*,\ee,\ee)+ \beta a \boldsymbol{D}^*(\rr-\rr_0^*,\ee)+\gamma a \boldsymbol{G_Q}^*(\rr-\rr_0^*,\ee,\ee)+\mathcal{O}(|\rr-\rr_0|^4),
\label{sing_wall}
\end{empheq}
where $r_0^*=(0,0,-h)$ is the image of the singularity located at $r_0$. Although the image of a Stokeslet has a simple expression and was computed long ago (see e.g. \cite{Blake1971}), the image of the singularities present in Eq. \eqref{sing_free}  (Stokeslet dipole, source dipole, Stokeslet quadrupole) have complicated expressions but are given in \cite{Spagnolie2012}:
 \begin{itemize}
  \item the image of the Stokeslet dipole $\boldsymbol{G_D}^*$ is given in \cite{Spagnolie2012} Eq. (B5),
  \item the image of the source dipole $\boldsymbol{D}^*$ is given in \cite{Spagnolie2012} Eq. (B13),
  \item the image of the Stokeslet quadrupole $\boldsymbol{G_Q}^*$ is given in \cite{Spagnolie2012} Eq. (B8).
\end{itemize}

In the main text, as examples of the influence of the solid wall on the calculated flow field, we plot on Fig. 5 from the main text two components of the flow field around the moving swimmer using the three different analytical expressions, obtained respectively with the singularities approximation in free space [Eq. \eqref{sing_free}], the singularities approximation with the influence of the wall [Eq. \eqref{sing_wall}] and the exact solution in free space valid at any point [Eq. \eqref{field_eq_moving}].

\section{Consistency between the measurements of the flow fields around a stuck colloid and around a moving colloid}

\begin{figure}
\begin{center}
\includegraphics[width=10cm]{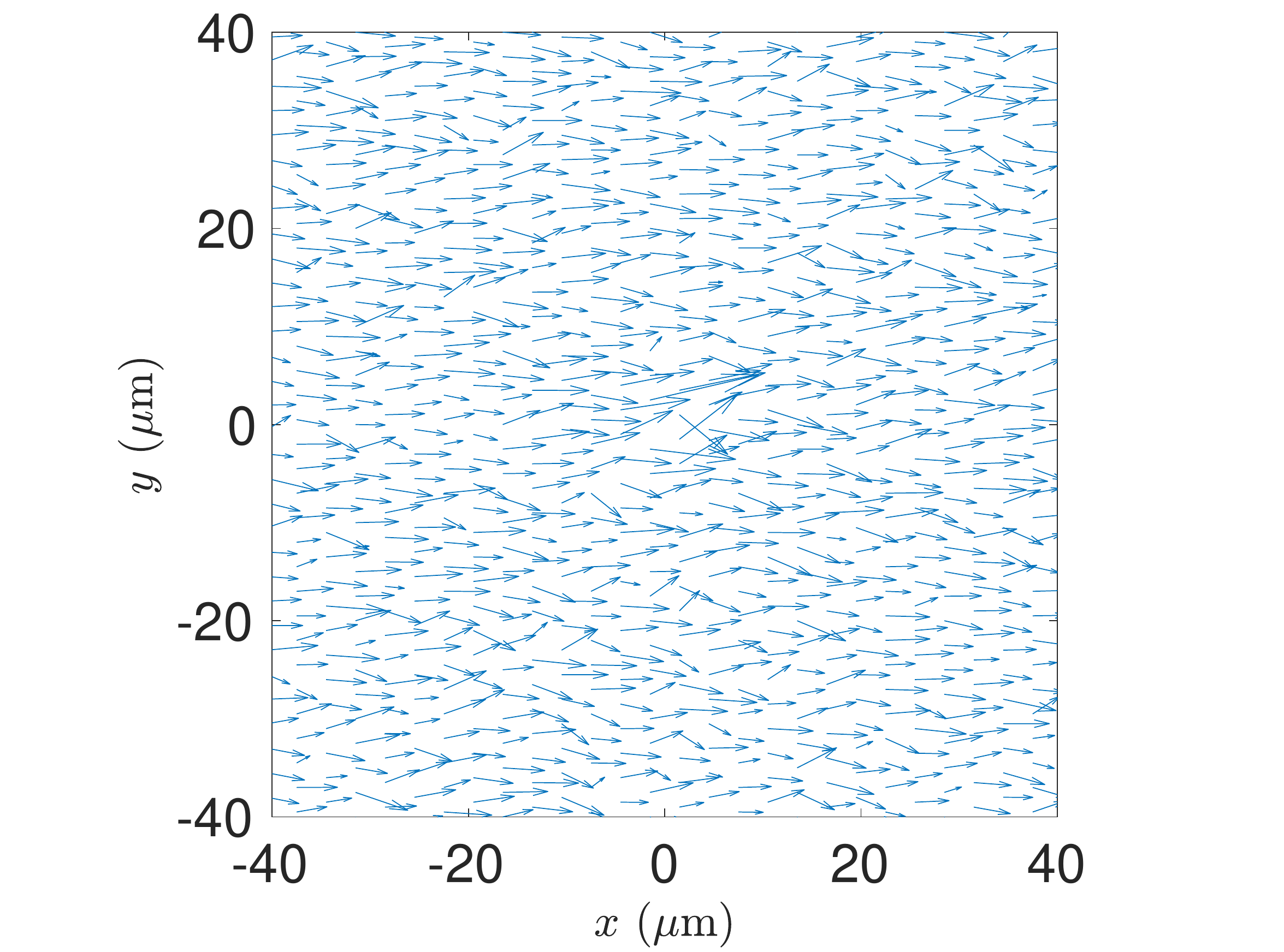}
\caption{Vector plot of $\delta \uu$ as defined in Eq. \eqref{ustuck_decompose}.}
\label{deltau}
\end{center}
\end{figure}

The velocity field around a stuck colloid can be related to the velocity field around a moving colloid through
\begin{equation}
\label{ustuck_decompose}
\uu^{\text{stuck}} = \uu^{\text{moving}}+U\ee + \uu^{\text{monopole}}
\end{equation}
where $\uu^{\text{stuck}} $ is given by Eq. \eqref{field_eq_stuck}, $\uu^{\text{moving}} $ is given by Eq. \eqref{field_eq_moving},  and $\uu^{\text{mono}} $ is given by Eq. \eqref{umono}. In order to highlight the consistency between the measurements in the stuck and in the freely swimming cases, we compute the quantity $\delta \uu=\uu^{\text{stuck}}- \uu^{\text{moving}}-\uu^{\text{monopole}}$. The value of the force is chosen such that $f=6\pi\eta a U$ (see section \ref{sec_stuck}). We present on Figure \ref{deltau} below the vector field $\delta \uu$. As $\delta \uu$ is expected to be equal to $U\ee$ anywhere around the colloid, we compute $\langle \delta u_x \rangle$  and $\langle \delta u_y \rangle$, where the averages run over all measurement points and find
\begin{eqnarray}
\langle \delta u_x \rangle & = & (1.14\pm0.29)U, \\
\langle \delta u_y \rangle & = & (-0.0011 \pm0.28)U,
\end{eqnarray}
which is a strong indication that Eq. \eqref{ustuck_decompose} holds experimentally, and which highlights the consistency between the two sets of experimental measurements. 

\section{Estimate of the coefficient $K$ and of the error on this estimate}

For each situation (moving colloid and stuck colloid), the experimental data consists in two sets of measurements  $\{\tilde{u}_x(x_i)\}_i$ et $\{\tilde{u}_y(y_i)\}_i$.  The objective is to fit each set of data respectively with the functions $K u_x(x_i,0,0) \equiv K w_{x,i}$ and $K u_y(0,y_i,0) \equiv K w_{y,i}$ (defined in Eq. \eqref{field_eq_stuck} if the colloid is stuck or Eq. \eqref{field_eq_moving} if the colloid is moving) with a single parameter $K$. This is done by a least-square method which yields
\begin{equation}
\label{ }
K = \frac{\sum_i \tilde{u}_x(x_i) w_{x,i}+\sum_i \tilde{u}_y(y_i) w_{y,i}}{\sum_i  {w_{x,i}}^2+\sum_i {w_{y,i}}^2}.
\end{equation}
The error is estimated as
\begin{equation}
\label{ }
\delta K = \sqrt{\frac{1}{n-1}}\sqrt{\frac{ \sum_i (\tilde{u}_x(x_i) -Kw_{x,i})^2+\sum_i (\tilde{u}_y(y_i) -Kw_{y,i})^2}{\sum_i(w_{x,i}-\overline{w_x})^2+\sum_i(w_{y,i}-\overline{w_y})^2}},
\end{equation}
where $n$ is the number of data points.

\section{Flow fields generated by alternative simple slip velocity profiles}

\subsection{Slip velocity with a dipolar symmetry}

The flow field generated by the nonzero slip velocity at the surface of the colloid can again be calculated Eqs. \eqref{field_eq_stuck} and \eqref{field_eq_moving}, but with another expression for the slip velocity at the surface of the colloid [see Eq. (2) 
in the main text].

\begin{figure}
\begin{center}
\includegraphics[width=12cm]{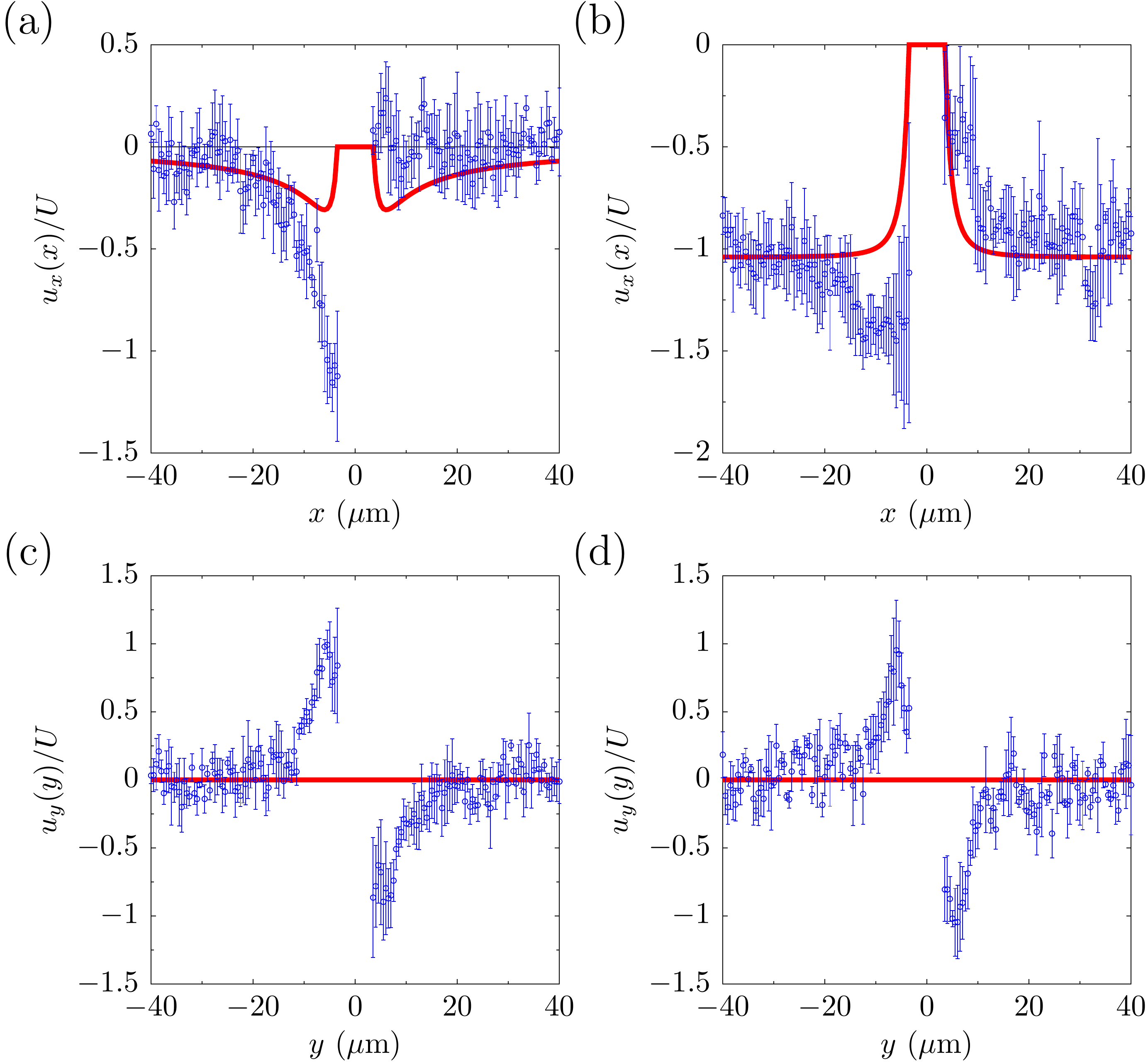}
\caption{Top: $x$-component of the velocity measured at $y=0$ and as a function of the coordinate $x$ for the two situations where the swimmer is stuck (a) and freely moving (b). Bottom: $y$-component of the velocity measured at $x=0$ and as a function of the coordinate $y$ for the two situations where the swimmer is stuck (c) and freely moving (d). The red line is a fit of the experimental data using a slip velocity with dipolar symmetry [Eq. \eqref{slip_dipolar_sym}]. The values of the fit parameters are  $K_\text{stuck}=0.80\pm0.26$ and $K_\text{moving}=1.56\pm0.34$.}
\label{other_slip_dipolar_sym}
\end{center}
\end{figure}

We first consider a slip velocity with a dipolar symmetry, that is peaked at the equator of the particle and vanishes at its poles:
\begin{equation}
\label{slip_dipolar_sym}
v(\theta) = v_0 \sin\theta 
\end{equation}
The velocity profiles $u_x(x,y=0,z=0)$ and $u_y(x=0,y,z=0)$ can be calculated and fitted to the experimental measurements using $v_0$ as a fit parameter (Fig. \ref{other_slip_dipolar_sym}). The results are commented in the main text.

\subsection{Constant slip velocity over the Pt hemisphere}

\begin{figure}
\begin{center}
\includegraphics[width=12cm]{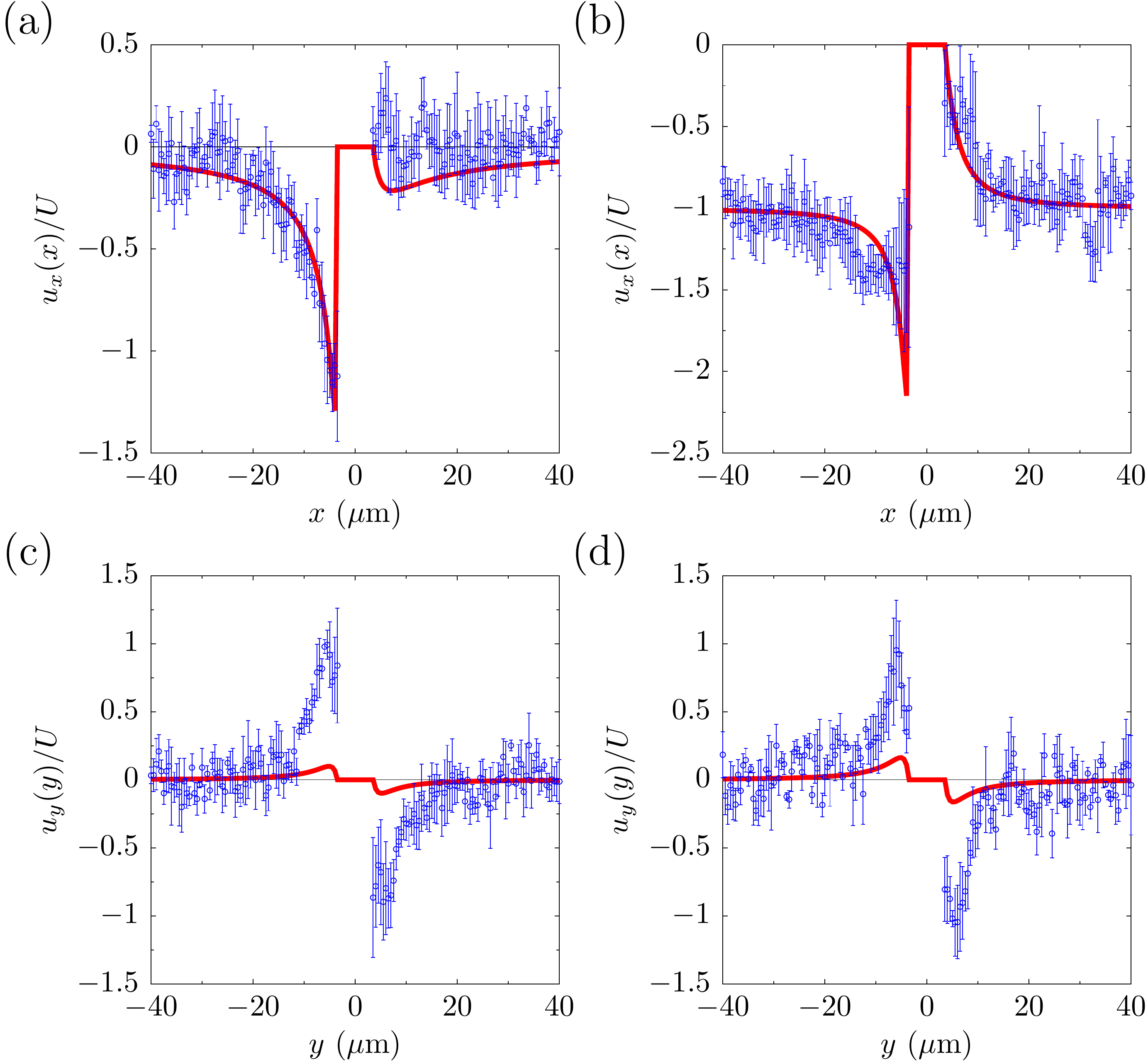}
\caption{Top: $x$-component of the velocity measured at $y=0$ and as a function of the coordinate $x$ for the two situations where the swimmer is stuck (a) and freely moving (b). Bottom: $y$-component of the velocity measured at $x=0$ and as a function of the coordinate $y$ for the two situations where the swimmer is stuck (c) and freely moving (d). The red line is a fit of the experimental data using constant slip velocity over the Pt hemisphere [Eq. \eqref{slip_constant}]. 
The values of the fit parameters are  $K_\text{stuck}=1.55\pm0.14$ and $K_\text{moving}=2.55\pm0.21$.}
\label{other_slip_constant}
\end{center}
\end{figure}

We then consider a slip velocity that is constant over the Pt hemisphere:
\begin{equation}
\label{slip_constant}
v(\theta) =
\begin{cases}
  v_0    & \text{ for $\pi/2<\theta<\pi$}, \\
  0    & \text{otherwise}.
\end{cases}
\end{equation}
Although this is not an existing prediction for the system of Pt-PS colloid, we believe its comparison with the model that is based on current loops will be helpful. The velocity profiles $u_x(x,y=0,z=0)$ and $u_y(x=0,y,z=0)$ can be calculated and fitted to the experimental measurements using $v_0$ as a fit parameter (Fig. \ref{other_slip_constant}). The results are commented in the main text.

\end{document}